\documentclass[letterpaper,twocolumn,10pt]{article}
\usepackage{usenix2019_v3}

\usepackage{tikz}
\usepackage{amsmath}
\usepackage{booktabs} % For formal tables
\usepackage{enumerate}
\usepackage{epsfig}
\usepackage{graphicx}
\usepackage{epstopdf}
\usepackage{amsmath}
\usepackage{amssymb}
\usepackage[linesnumbered,ruled]{algorithm2e}
\usepackage{setspace}
\usepackage{url}
\usepackage{listings}
\usepackage{subfigure}
\usepackage{rotating}
\usepackage{lscape}
\usepackage{verbatim}
\usepackage{hhline}
\usepackage{textcomp,booktabs}
\usepackage{xcolor}
\usepackage{paralist}
\usepackage{hyperref}
\usepackage{grffile}
\usepackage{mathtools}
\usepackage{amsthm}
\usepackage{hyphenat}
\usepackage{datatool}
\usepackage{numprint}
\usepackage{color, colortbl}
\usepackage{flushend}
\usepackage{enumitem}
\usepackage{xurl}
\usepackage[autostyle,threshold=0]{csquotes}
\usepackage{datatool}

\npthousandsep{,}

\DTLsetseparator{;}
\DTLloaddb{expdata}{figures/data/experimental_data.txt}
\DTLforeach*
{expdata}% database label
{\today=today, \timelinescount=timelinescount,\trendscount=trendscount,\actagemin=actagemin, \actagemax=actagemax, \actagemean=actagemean, \actagesd=actagesd, \hashmonitor=hashmonitor}% assignment

\makeatletter
\newcommand{\removelatexerror}{\let\@latex@error\@gobble}
\makeatother

\let\OLDthebibliography\thebibliography
\renewcommand\thebibliography[1]{
  \OLDthebibliography{#1}
  \setlength{\parskip}{0pt}
  \setlength{\itemsep}{0pt plus 0.3ex}
}

\begin{document}
\begin{filecontents*}{figures/data/experimental_data.txt}
today;timelinescount;trendscount;actagemin;actagemax;actagemean;actagesd;hashmonitor
April 28, 2021;264043;5499700;11.86;139.01;93.14;48.82;3.1
\end{filecontents*}

\date{}

\title{\Large \bf Strategies and Vulnerabilities of Participants in Venezuelan Influence Operations}
\author{
{\rm Ruben Recabarren \qquad Bogdan Carbunar \qquad Nestor Hernandez \qquad Ashfaq Ali Shafin}\\
FIU}
\maketitle

\begin{abstract}
Studies of online influence operations, coordinated efforts to disseminate and amplify disinformation, focus on forensic analysis of social networks or of publicly available datasets of trolls and bot accounts. However, little is known about the experiences and challenges of human participants in influence operations. We conducted semi-structured interviews with 19 influence operations participants that contribute to the online image of Venezuela, to understand their incentives, capabilities, and strategies to promote content while evading detection. To validate a subset of their answers, we performed a quantitative investigation using data collected over almost four months, from Twitter accounts they control.

We found diverse participants that include pro-government and opposition supporters, operatives and grassroots campaigners, and sockpuppet account owners and real users. While pro-government and opposition participants have similar goals and promotion strategies, they differ in their motivation, organization, adversaries and detection avoidance strategies. We report the Patria framework, a government platform for operatives to log activities and receive benefits. We systematize participant strategies to promote political content, and to evade and recover from Twitter penalties. We identify vulnerability points associated with these strategies, and suggest more nuanced defenses against influence operations.
\end{abstract}

\section{Introduction}

Social networks have become the central medium for {\it influence operations} (IOs), enabling them to disseminate and amplify disinformation, and compromise the integrity of information posted by others. We observe a parallel between disinformation (information designed to mislead~\cite{SAW19}) and malware (e.g., self-propagating worms) where disinformation runs on human minds instead of computers. From this perspective, influence operations seek to fraudulently boost the search rank of the content they distribute, and increase the number of human hosts exposed and infected.

Goals of influence operations include manipulating or corrupting public debate, undermining trust in democratic processes and scientific evidence, and even influencing elections~\cite{FacebookIO, ABST20, BloombergGuide, Journalists.Trolls}. Such efforts are becoming increasingly prevalent: Facebook reported the discovery of 150 covert IOs on its site between 2017 - 2020, that originated from countries all over the world~\cite{FacebookIO}.

\begin{figure}
\centering
\includegraphics[width=0.95\columnwidth]{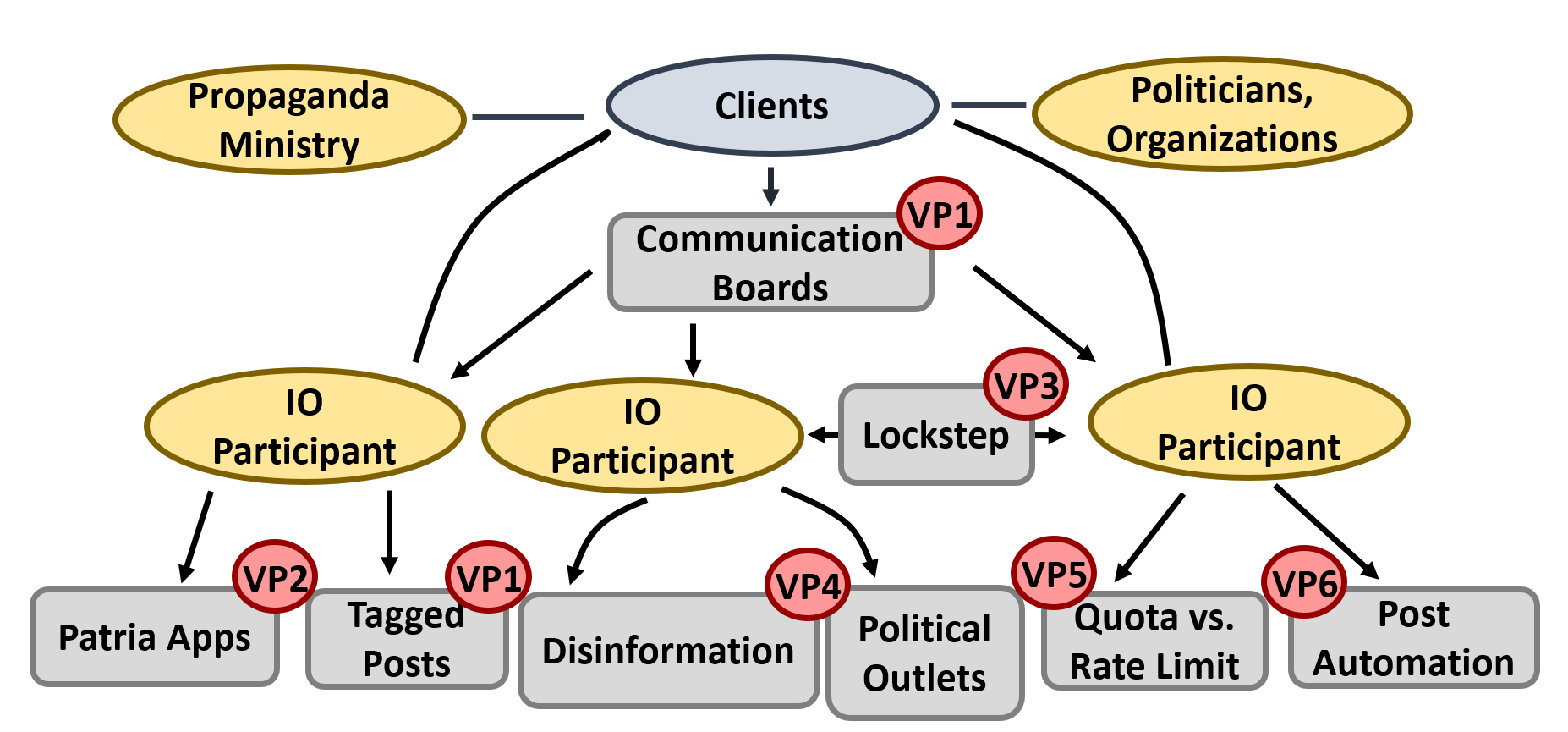}
%\vspace{-10pt}
\caption{Map of discovered strategies (gray rectangles) of influence operations participants (yellow ovals). Red circles represent influence operations vulnerabilities that we identified, and discuss in the context of Twitter changes, in $\S$~\ref{sec:assurance}.}
%\vspace{-5pt}
\label{fig:strategies}
\end{figure}

Influence operations were shown to be well organized~\cite{SAW19, BloombergGuide, IRA, NYT.Trolls, LGF20, China.army, KCLS17}, control many social network sockpuppet accounts~\cite{BloombergGuide, LW18, KCLS17}, and employ inauthentic behaviors~\cite{FacebookIO, Russia.Shift, ZCSSSB19, NYT.Trolls, TrollsNow}. This knowledge was collected through journalistic efforts~\cite{BloombergGuide, USSpanishCampaigns, Russia.Shift, Russia.division, TrollsNow, Journalists.Trolls, NYT.Trolls} and forensic analysis of social networks~\cite{KPR17, LW18, ABLF19, ICSLDBHJG20, ZCDSSB19, ZCSSSB19} and released datasets~\cite{SAW19, Twitter.Data, Troll.Tweets, Troll.Tweets2, Facebook.Ads, RedditIRA}.

To develop information assurance solutions that can control influence operations, we need however to understand the experiences, challenges and vulnerabilities of their contributors. We lack such information due to difficulties to identify, reach, recruit and establish trust with such participants.

In this paper, we investigate the perspective of participants in influence operations. For this, we leverage unique background and insights into Venezuela, a country where influence operations have replaced verified news~\cite{BloombergGuide, BloombergDocument, NewsBattle, InternetGrip}. Since 2013, Maduro's regime has taken over the country's institutions, electoral and justice system, and has censored standard news delivery solutions~\cite{VenezuelaCensorship}. To bypass censorship, communicate and organize, the opposition uses social networks and mobile apps~\cite{HZ19}; conversely, the government uses them to distribute hyperpartisan news and disinformation~\cite{BloombergGuide}.

In a first contribution, we developed a protocol to identify and recruit participants in Venezuelan influence operations. Our protocol uses Telegram groups and Twitter to identify candidates, and to contact them over direct messaging. Second, we recruited 19 relevant participants, and conducted semi-structured interviews to study the following key questions:

\begin{compactitem}

\item
{\bf RQ1}: What are concrete (a) organization and communication mechanisms, (b) resources, capabilities, and limitations, (c) motivation, and (d) promotion strategies of participants in Venezuelan influence operations?

\item
{\bf RQ2}:
Do they participate in influence operations that target other countries? (a) Are they willing to be hired to participate in external influence operations?

\item
{\bf RQ3}:
What is the participant perception on disinformation? (a) Do they contribute or do they have strategies to avoid their distribution?

\item
{\bf RQ4}: Are participants aware of, and affected by social network defenses and penalties? (a) Have they developed strategies to circumvent and recover from detection? (b) Are these strategies effective?

\end{compactitem}

Third, to validate participant claims, we performed a quantitative investigation with data collected over four months, from 34 Twitter accounts they control.
%
%We further place our results in the context of 2.9 million Twitter posts that we collected from a hyperactive, hyperpartisan subset of the 314,646 followers of a key influence operations source revealed by our participants.
%
Our findings include:

(1) Interview participants are diverse, e.g., pro-government vs. opposition supporters, paid operatives vs. grassroots campaigners, and sockpuppet owners vs. real users ($\S$~\ref{sec:results:capabilities}). We found consistency with the ``communication constitutes organization'' perspective of organizational theory~\cite{PN09} ($\S$~\ref{sec:results:organization});

(2) Both pro-government and opposition participants revealed a history of contribution to foreign campaigns. Many on both sides are willing to be hired to participate in influence operations, including targeting US politics ($\S$~\ref{sec:results:international});

(3) Participants claimed strategies to verify information they post. However, we report concrete instances of pro-government participant distribution of disinformation that received significant community engagement ($\S$~\ref{sec:results:fakenews});

(4) Adversarial environments: Pro-government participants reported efforts by social nets to thwart their activities; the opposition revealed pro-government operative attacks against their accounts ($\S$~\ref{sec:results:defenses}). Both sides disclosed strategies to avoid detection and recover from account suspensions ($\S$~\ref{sec:results:avoidance});

(5) Pro-government and opposition participants differ in their motivation, organization, adversaries, and detection avoidance strategies. Based on our findings, we present the IO strategy map of Figure~\ref{fig:strategies} ($\S$~\ref{sec:results:strategies}).

In a fourth contribution, we identify vulnerability points (VPs) associated with promotion strategies revealed by our participants (Figure~\ref{fig:strategies}), and suggest changes to social networks' handling of influence operations ($\S$~\ref{sec:assurance}).

\section{Background, Model and Goals}
\label{sec:background}

\subsection{The Venezuelan Crisis}

Venezuela is experiencing the worst economic crisis in its history~\cite{VenezuelaCrisis}. The government controls every aspect of daily life ranging from food to gas supply, while the country struggles with hyperinflation, unemployment and poverty. In recent years, Maduro's government has implemented a takeover of the Venezuelan government and institutions, the electoral and justice systems, and the army. The government has used lethal force against protesters, exiled critics, and held political prisoners. Six million people have migrated to neighboring countries~\cite{S20}.

A large majority of Venezuelans oppose the current regime~\cite{MaduroExit}. The government has however invested heavily in media censorship efforts~\cite{VenezuelaCensorship}. This has led to a migration of anti-government movements to social media and mobile apps~\cite{NewsBattle, InternetGrip}. In turn, this was followed by the creation of a government-sponsored online army~\cite{BloombergDocument}, to disrupt the opposition activities and promote the government propaganda.

Political allies of Venezuela provide support, by distributing hyperpartisan news and disinformation through their Spanish language news organizations~\cite{VenezuelaRussia, FabFive}, e.g., Russia Today (RT) Español~\cite{ActualidadRT}, Sputnik Mundo~\cite{SputnikMundo}, the Iranian Hispan TV~\cite{HispanTV} and Cuban~\cite{CubaInformacion, CubaSi} news outlets.

\subsection{Adversary}
\label{sec:background:adversary}

\noindent
{\bf Influence Operations}.
Influence operations (IOs) also known as information campaigns~\cite{TABBBCDDKKMMRS21} or strategic information operations~\cite{SAW19}, are coordinated efforts to manipulate or corrupt public debate for a strategic goal~\cite{FacebookIO}. Influence operations were shown to have at least short-term effects, that include political beliefs and behavior changes~\cite{BABBCHLMMV18}, increased xenophobia~\cite{WBJLO20}, and increased uncertainty about vaccines~\cite{CJA20}.

%While most organizations conduct influence operations to promote their own agenda, services that provide IOs to customers, with deniability, are also emerging~\cite{FacebookIO, OC19}.

In this work we distinguish between centralized influence operations and grassroots movements. While grassroots movements are bottom-up, often spontaneous decision making efforts~\cite{Y17}, centralized influence operations have a command and control (C\&C) center (e.g., government, institution, or interest group) that designs the operation's goals and message.

We now define several types of participants in influence operations. Not all are adversarial. We discuss them here because they are used or manipulated by adversarial C\&Cs.

\noindent
{\bf IO Participants}.
To avoid detection, centralized operations were shown to emulate online grassroots movements~\cite{FacebookIO, OC19}. They achieve this by recruiting real people, that include {\it operatives} and {\it grassroots campaigners}. Operatives receive incentives to promote the operation's message online~\cite{FacebookIO, OC19}; grassroots campaigners believe the information they distribute, do not get paid and are unwilling to be hired for activities that contradict their beliefs. Our study includes both types of participants. Such participants were shown to create and amplify posts that promote the operation's message and to communicate and coordinate activities~\cite{SAW19}.

In contrast, {\it unwitting agents}~\cite{B85, SAW19} are human participants that receive and occasionally engage with influence operations content, but do not receive external incentives and do not coordinate activities.

Influence operations contributors also include {\it trolls}, that use anonymous accounts and post inflammatory and digressive messages, designed to trigger conflict and disrupt online discussions~\cite{BloombergGuide, march2019psychopathy}. Datasets of Russia's Internet Research Agency (IRA) trolls in Twitter~\cite{Twitter.Data, congressIRA} revealed several types of troll accounts, each performing a specialized function~\cite{LW18}. While IRA tweets reached many users~\cite{ZCDSSB19}, they had minor impact in making content viral~\cite{ZCDSSB19}. In contrast, we found that many of our participants accumulated significant community engagement for their posts, including disinformation.

Influence operations also use {\it bots}, automated accounts that require little human supervision~\cite{JHHPOWV20, AYM15, FVDMF16}. Both trolls and bots use sockpuppet accounts to hide their identities. Previous work however has found that many IO participants are not bots, and manage their online identities in complex ways~\cite{ICSLDBHJG20}. Most of our interview participants use their own identity to establish a personal brand and a follower base.

\noindent
{\bf Coordination Apps}.
IO participants use apps to communicate and coordinate activities~\cite{SAW19}. Previous work studied misinformation and fear speech in WhatsApp~\cite{JUIQTCG22,kazemi2021tiplines}. In particular, Javed et al.~\cite{JUIQTCG22} analyzed the spread of information through WhatsApp, and documented information flows between WhatsApp and Twitter. Our participant recruitment process builds on a similar finding, that Venezuelan operators use communication apps to organize, coordinate and disseminate messages to be promoted in Twitter ($\S$~\ref{sec:methods}). We also found similar information flows in Venezuelan operations, between Telegram groups and Twitter.

\subsection{Research Goals}
\label{sec:background:goals}

Social networks implement various techniques to address influence operations. They include mechanisms to detect~\cite{XGCKJLSPFP21}, verify~\cite{KYKBFPI20} and penalize accounts and activities that violate their terms of service. Twitter penalties include suspending accounts detected to post spam, suspected to be compromised, or reported to violate rules surrounding abuse~\cite{TwitterSuspension}. Further, social networks were reported to {\it shadowban}, i.e., remove or limit the distribution or visibility of certain content~\cite{TwitterToS, ShadowCensorship}.

Such mechanisms are often unable to address influence operations in real time~\cite{FacebookIO}, and some consider them to be censorship~\cite{ShadowCensorship}. Our study confirms this.

Instead, in this work we seek to provide insights into influence operations, by studying the perspective of IO participants. We document experiences, motivation, organization and communication mechanisms, capabilities, goals and strategies of participants in Venezuelan influence operations, in order to understand their strengths and vulnerabilities, and help inform future efforts to design more appropriate, inclusive and effective solutions to address influence operations.

\section{Methodology}
\label{sec:methods}

Our study consists of a qualitative exploration and a quantitative investigation into various aspects of participation in influence operations. We first detail the recruitment procedure and ethical considerations, then describe the studies.

\subsection{Participant Recruitment}

\begin{figure}[t]
\begin{minipage}{0.97\columnwidth}
\removelatexerror% Nullify \@latex@error
\begin{algorithm}[H]
\begin{tabbing}
XXX\=X\=X\=X\=X\=X\= \kill
1.{\mbox{\bf{StudyProtocol}}(SN: SocialNet, IOGroups: list)}\\
2.\>{$\mathtt{\mbox{\bf{while}}\ (true)\ \mbox{\bf{do}}\{}$}\\
3.\>\>{$\mathtt{IOMembers = getSNAccounts(IOGroups);}$}\\
4.\>\>{$\mathtt{IOActive = followBack(IOMembers);}$}\\
5.\>\>{$\mathtt{IOFollowers = getFollowers(IOActive);}$}\\
6.\>\>{$\mathtt{candidates = getOpenToDM(IOFollowers);}$}\\
7.\>\>{$\mathtt{respondents = sendDM(candidates);}$}\\
8.\>\>{$\mathtt{groups = \{\};}$}\\
9.\>\>{$\mathtt{\mbox{\bf{for\ each}}\ R\ \mbox{\bf{in}}\ respondents}$}\\
10.\>\>\>{$\mathtt{(answers, accounts, g) = interview(R);}$}\\
11.\>\>\>{$\mathtt{accountData = SN.collectData(accounts);}$}\\
12.\>\>\>{$\mathtt{validate(answers, g, accountData);}$}\\
13.\>\>\>{$\mathtt{groups = groups\ \cup\ g;}$}\\
14.\>\>{$\mathtt{\mbox{\bf{if}}\ (groups \in IOGroups)\ \mbox{\bf{then}}\ break;}$}\\
15.\>\>{$\mathtt{IOGroups = IOGroups\ \cup\ groups;}\}$}
%\vspace{-75pt}
\end{tabbing}
\caption{Pseudocode for study protocol. The recruitment takes place on a social network SN (Twitter in our case). IOGroups lists influence operations communication groups used to seed the candidate search (from Telegram in our case).}
\label{alg:recruitment}
\end{algorithm}
\end{minipage}
%\vspace{-15pt}
\end{figure}

We focused recruitment efforts on identifying Twitter accounts with a verifiable history of participation in influence operations. Our recruitment protocol identifies active operatives, by starting with a seed set of communication groups. To identify this set, we leveraged observations that Venezuelan operatives use Telegram to communicate about their goals and objectives. We have used Telegram's search (keyword ``Twiteros activos'') to identify three Telegram groups and channels dedicated to Venezuelan influence operations.

Members of these groups often disclose their Twitter handles to follow one another. We have selected a set of Twitter accounts that were revealed by members of these groups. We did not contact these accounts directly: To minimize the chance of interference with our study, we wanted to delay news of our efforts from reaching the influence operations command and control center ($\S$~\ref{sec:background:adversary}). Members of these Telegram groups may have communication channels with the command and control, thus may quickly alert many participants and influence their perception about our study.

Instead, we followed these accounts from our lab's Twitter account. For the accounts that followed us back, we collected, via breadth-first search, and using the Twitter API, their Twitter followers. From these, we identified the accounts that were open to direct messaging from our account.

We sent an interview invitation to these accounts, over direct messaging (DM). We then sent personalized messages, including a consent form, to the accounts that replied. We inspected the accounts that accepted the consent form, and interviewed those that were active, were posting tweets with political topics and had at least 500 followers.

During the interview, we also collected other accounts claimed to be controlled by participants, and groups they claimed to use for communications. We then iterated our recruitment activities over these groups.

Algorithm~\ref{alg:recruitment} summarizes our study procedure, including recruitment, interviews, and data validation steps.

In total, we followed 1,543 Twitter accounts. From the 109 accounts that followed us back, we collected their 256,770 followers. We sent DMs to the subset of 2,843 accounts that were open for DM, then sent personalized messages to the 99 accounts that replied. From the 35 Twitter accounts that accepted the consent form, we selected 19 for interview.

\subsection{Ethical Considerations}
\label{sec:methods:ethical}

The study procedure was scrutinized and the full study was approved by the institutional review board of our university (IRB-20-0550). We followed ethical practices for conducting sensitive research with vulnerable populations~\cite{BBMR17}. For instance, we tailored the consent process to the participant, and re-confirmed consent. We sent the consent form link and obtained consent both during recruitment, and at the beginning of the interview. We obtained consent both electronically and verbally. We accommodated participant requests for private payments: cryptocurrencies, intermediaries in Venezuela, and sending money over snail mail.

During recruitment and the interview, we clearly declared the identity of the researchers, the research objective, the data that we collect (including Twitter account data) and how we process it, and potential impact on the participant. More specifically, our invitation and consent form made clear that our intention is to study political content promotion capabilities, resources and behaviors on social platforms. During recruitment, we followed candidate accounts from our lab's Twitter account, where we made clear its association to our lab, and its use strictly for research purposes.

We also explained any risks that their work may have through our research. We asked several times during the interview if they are comfortable discussing potentially sensitive topics and told them that they could skip any question.

We were careful to hide participant identity. Following the account data collection and participant payment steps, we removed all participant PIIs from our data (e.g., names, IDs, handles, locations). From the participants' Twitter accounts, we kept only account statistics, their posts, and their followers. We used multiple solutions to securely store de-identified research data. The data was stored on a physically secure Linux server in our university, and accessed through encrypted channels only from the password-protected authors' laptops. Further, all data was processed only on the server.

In $\S$~\ref{sec:discussion} we further revisit ethical considerations from the perspective of the impact of our findings.

\noindent
{\bf Team Positionality Statement}.
The research team consists of Venezuelan and international investigators, all located outside Venezuela. The investigators support neither the Venezuelan government or the opposition. The interviews were conducted by a politically neutral team member, who shared context with study participants, including the language and some knowledge of the country's political and economical circumstances.

\subsection{Qualitative Study}
\label{sec:methods:qualitative}

\noindent
{\bf Interviews}.
We conducted semi-structured interviews with the recruited participants: one pilot interview to test our interview guide and method, then in-depth interviews with 19 participants. The interview focused on participant (1) incentives to contribute to influence operations, (2) organization structure, (3) resources, capabilities and limitations, (4) strategies employed to promote IO goals, (5) operations in which they participated and in which they are interested to participate, (6) perception of disinformation in influence operations, (7) perception on the impact of Twitter's defenses on IO activities, and (8) strategies to evade and recover from detection.

%All participants were asked questions in the same order.

The interviews were conducted over the phone, in Spanish, by one author who is a native speaker. All audio interviews were recorded with participant permission. The interviews lasted between 17 and 98 minutes (M = 52, SD = 19.43). We paid 2.5 USD for every 15 minutes spent in the interview.

\noindent
{\bf Analysis Process}.
We analyzed responses using a grounded-theory open-coding process~\cite{SC97}, performed by two co-authors: the one who conducted the interviews and a non-Spanish speaker. We conducted the interviews over 6 weeks. During this time we also transcribed and anonymized the recorded interviews, then translated them into English. Given time constraints, we started the analysis after data collection was complete. Following each interview, the interviewer discussed impressions, observations and findings with the rest of the team. This enabled us to detect reaching data saturation~\cite{GBJ06}, where the interviewer reported no new insights from the last two participants. We confirmed this during analysis.

In the preliminary analysis stage, we independently read five transcripts to establish a thematic framework of the interview data. We coded participant responses to each interview question including relevant information provided later in the interview. We organized the themes into an initial codebook. We then independently coded and met to revise the codebook. We used these themes to organize codes emerging from the remaining 14 transcripts. Two co-authors met to discuss the themes and codes after processing each set of two to three interviews. In total, we created 177 codes from 410 pages of transcripts. Since we reviewed the coded transcripts jointly, we do not include the inter-rater reliability score~\cite{MSF19}.

\subsection{Quantitative Investigation}
\label{sec:methods:quantitative}

To validate participant claims, we performed a quantitative analysis with data from several sources:

\noindent
{\bf Participant Twitter Accounts}.
We have collected information from 19 Twitter accounts that we know are controlled by the participants, i.e., they replied to DMs we sent to these accounts during recruitment. We call these {\it recruitment validated} accounts. The accounts were between 11 months and 11.5 years old (M=\actagemean\ months, SD=\actagesd). We have monitored the Twitter timelines of these accounts over four months in 2021. We have collected their tweets and retweets, the engagement received by each tweet, the number of followers and accounts that they follow. We have collected the trending hashtags for all nine Venezuelan regions available in Twitter during that interval. In total, we have collected \numprint{\timelinescount} timeline posts and \numprint{\trendscount} trending hashtag reports.

In addition, 11 participants revealed during interviews, 15 other Twitter accounts they claimed to control.

\noindent
{\bf Telegram Groups}.
During participant recruitment, we have identified and joined six Telegram groups (TuiterosDeChavez, Tuiteros Patriotas, TuiterosActivos, Twiteros Patriotas, Twiteros Activos, Bonos de la Patria) used by participants to communicate and coordinate activities. The groups had a total of 3,352 members, and were active at the time of submission. The groups provide members with instructions regarding the work they are expected to perform.

\section{Results}
\label{sec:results}

In this section we first classify the participants, then explore perception and participation in the distribution of disinformation ($\S$~\ref{sec:results:fakenews}), motivation ($\S$~\ref{sec:results:motivation}), and willingness to participate in paid campaigns ($\S$~\ref{sec:results:international}). We then describe participant reported organization and communication channels ($\S$~\ref{sec:results:organization}), and capabilities ($\S$~\ref{sec:results:capabilities}). We discuss reported strategies to promote content ($\S$~\ref{sec:results:strategies}), perceptions of Twitter defenses ($\S$~\ref{sec:results:defenses}), and strategies to evade and recover from detection ($\S$~\ref{sec:results:avoidance}).

\subsection{Participant Classification}
\label{sec:results:participants}

\begin{figure}
\centering
\includegraphics[width=0.79\columnwidth]{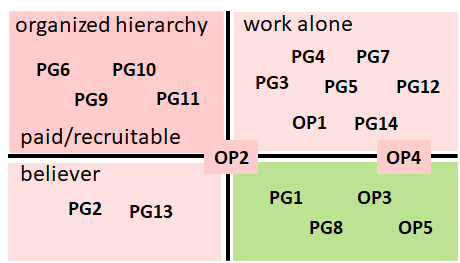}
%\vspace{-10pt}
\caption{Participant classification across two dimensions: (1) member of organized hierarchy (left column) vs. working alone (right column), and (2) paid or willing to be hired (top row) vs. believer (bottom row). Our participants include both influence operators and grassroots members.}
\label{fig:participant:classification}
%\vspace{-5pt}
\end{figure}

\noindent
{\bf Demographics}.
Our participants have diverse backgrounds. Thirteen male, seven female; age range between 18 and 67 (M=50.8, SD=11.5); job types include self-employed (2), teacher (7), engineer (2), lawyer (1), public accountant (2), communications expert (2), manager (1), TV actor (1) and assistant (2). The highest education level was high-school (4), bachelors (10), masters (5) and PhD (1). 18 participants lived in Venezuela, one in Nicaragua.

\noindent
{\bf Pro-government vs. Opposition}.
Fourteen participants were pro-government and five supported the opposition. We verified this using their Twitter account data. In the following, for simplicity, we use PG1, .., PG14 for the pro-government participants and OP1, .., OP5 for the opposition participants.

\noindent
{\bf Operatives vs. Grassroots Campaigners}.
Figure~\ref{fig:participant:classification} shows the classification of our participants on two dimensions: (1) members of an organization vs. working alone, and (2) having received benefits or being willing to be hired vs. being a believer. Six pro-government and one opposition participant operated in hierarchical operations, and received or issued instructions. Overall, twelve pro-government and one opposition participants were either part of an organized hierarchy, had received rewards, or were willing to receive rewards for their activities.

Two pro-government and two opposition participants have strong political convictions, and may be considered grassroots campaigners ($\S$~\ref{sec:background}). Two other opposition participants are hybrid (OP2, OP4, shown on borderlines in Figure~\ref{fig:participant:classification}). For instance, OP2 used to be pro-government, and had leadership roles in influence operations. OP2 later became disillusioned, started supporting the opposition, and was even a political prisoner. OP2's activities are driven by political beliefs, but is also recruited to participate in campaigns during special events, e.g., before elections.

\subsection{RQ3: Disinformation}
\label{sec:results:fakenews}

% \begin{figure}
% \centering
% \includegraphics[width=0.99\columnwidth]{./figures/tweets/tweets.png}
% \vspace{-10pt}
% \caption{Example tweets (translated from Spanish) posted by interview participants.}
% \label{fig:tweets:examples}
% \end{figure}

% \begin{figure}
% \centering
% \includegraphics[width=0.89\columnwidth]{./figures/tweets/works.png}
% \vspace{-10pt}
% \caption{Sample image collage promoted by pro-government participants to advertise government efforts to improve the situation in Venezuela.}
% \label{fig:tweets:works}
% \end{figure}

\noindent
{\bf The Twitter Truth}.
All participants have created or amplified hyperpartisan news in Twitter.
%
%Figure~\ref{fig:tweets:examples} shows sample tweets posted by opposition and pro-government participants.
%
We observed consistency among the views of pro-government participants, who often re-tweeted the same posts. This includes images from staged events, showing efforts by various institutions and politicians to improve the lives of Venezuelans.
%
%(see Figure~\ref{fig:tweets:works}).
%
Given the government's obliteration of independent reporting, such events are impossible to verify, and highly suspicious. We observed more diverse interests among opposition participants. However, they also post and promote anti-government messages and accusations, often without providing trustworthy proofs.
%
%(see Figure~\ref{fig:tweets:examples}).

\begin{figure}
\centering
\includegraphics[width=0.97\columnwidth]{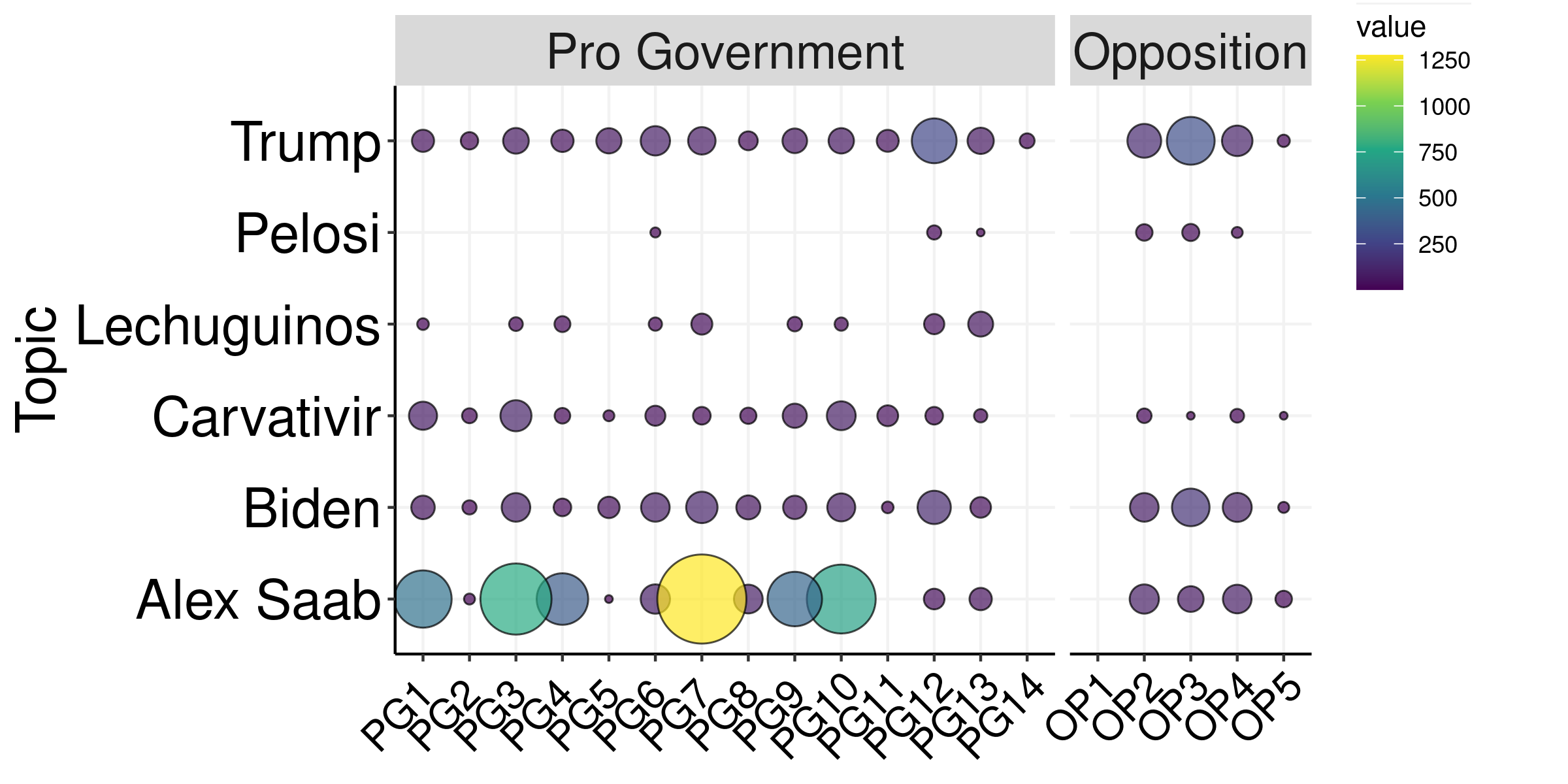}
%\vspace{-15pt}
\caption{Per-participant number of posts over four months, on select controversial and US politics-related topics. Both pro-government and opposition participants have interests in US politics and controversial topics, but with opposing views.}
\label{fig:interests}
%\vspace{-5pt}
\end{figure}

Figure~\ref{fig:interests} shows the number of posts from participants, on controversial subjects ``Carvativir''~\cite{Carvativir} (540 posts), ``Alex Saab''~\cite{AlexSaab} (4,577 posts) and articles from disinformation site~\cite{Lechuguinos} \url{lechuguinos.com} (72 tweets, 178 retweets from 9 participants). Carvativir is a thyme derivate that was promoted by the Venezuelan president to neutralize COVID-19 with no side effects, a claim not substantiated by data~\cite{Carvativir}. Alex Saab is a Colombian businessman, alleged financier for Venezuela's president, who was arrested and extradited to the US. He was accused by the US Department of Treasury to be part of the corruption network that stole from Venezuela’s food distribution program~\cite{TreasurySaab}, see also $\S$~\ref{sec:results:motivation}.

We observed however opposing views between these groups. For instance, on Carvativir~\cite{Carvativir}, pro-government participants distributed claims that FDA considers it to be safe, that it is optimal for the treatment of COVID-19, and has antiviral capacity to block SARS-CoV-2 and positive effects in COVID-19 patients. The opposition participants claimed that the government deceives people and uses Carvativir as a source of revenue. Further, while pro-government and opposition participants converge in their enmity toward the US president Biden, their reasons differ: Pro-government operations use him as a scapegoat to blame for the country's situation; opposition participants believe that his government will convert the US into Venezuela.

We further found 559 posts with links to Venezuelan government sites; also, 219 posts with links to Russian~\cite{ActualidadRT, SputnikMundo}, Iranian~\cite{HispanTV} and Cuban~\cite{CubaInformacion, CubaSi} news outlets, known to distribute disinformation~\cite{DWPropaganda, RussiaDisinformation}. This is consistent with strategies of integration of government and externally-funded media as source content for narratives in countries like Syria~\cite{SAWVYS18, SAW19}.

These findings confirm a ``firehose of propaganda and falsehood'' model~\cite{PM16} employed by pro-government participants, where propaganda and disinformation is used to drown out the opposition~\cite{OC19} and reduce the ability of readers to make sense of information~\cite{PW14, PM16}.

\noindent
{\bf Perception of Disinformation}.
Both pro-government and opposition participants explained that they have witnessed disinformation in Twitter, e.g., ``{There is a lot of fake news}'' (OP3), ``{\it Many people tend to post fake news}'' (PG5). To avoid distributing such posts, some explained that they research the content they receive, e.g., ``{\it I research [my publications] otherwise I could become an amplifier of what is known as fake news.}'' (PG7), ``{\it many times, we learn about something then we research the truth}'' (PG5). We emphasize however the lack of trustworthy news sources, the remote nature of many reported events, and restricted communications.
%
%(Appendix~\ref{appendix:findings:limitations}).

Some participants validate the sources of tweets, e.g., {\it
``I retweet posts from journalists and politicians that publish truthful information, and not accounts with pseudonymous and unknown names''} (OP3). This is consistent with findings that people in the US rate mainstream sources more trustworthy than hyperpartisan or disinformation sources~\cite{PR19}. However, others found that the source has little impact on how people judge headlines (accurate vs. inaccurate)~\cite{DPR20}.

Several participants claimed that when posting original tweets, they add links to a credible source that confirms the information. 11,678 of the 237,978 posts we collected from the accounts of our participants contained links to other sites.

%Two participants claimed that when they cannot provide evidence, they specify that the posted information is rumor or personal opinion.

\subsection{RQ1: Motivation}
\label{sec:results:motivation}

Twelve participants claimed to have received some form of rewards for their online activities. Of these, nine were not required to do this work, while three reported mechanisms suggesting coercion. Of the latter, one received medical help from the government, and two explained they are on government payroll, where posting political content is part of their work. Indeed, working for the government, which for many who cannot leave the country is the only option, entails being subject to implicit forms of both blackmail and bribery. For instance, state employees who do not tweet in favor of the government or who do not go to government-sponsored protests do not get paid or do not receive food stamps. We note that 60\% of the active population is employed, of which almost 30\% are working for the government~\cite{ElPaisEmployment}.

One theme among pro-government participants was the central role played by the government-commissioned Patria platform~\cite{Patria}, in recording online activities and distributing rewards. Patria was inspired by the Chinese social credit system, was developed by ZTE~\cite{PatriZTE}, and uses the Homeland ID Card to identify and link users across plans. The platform includes the Android vePatria app~\cite{vePatria}, the veMonedero app to connect the user wallet and receive bonuses, and the veQR app to keep track of social plans offered by the government.

Participants revealed that the Patria system provides (1) {\it activity awards}, for accounts that post around 50 tweets and at least 300 retweets a day, (2) a bonus if their posts receive significant engagement from other participants, and (3) monthly bonuses through the Carnet de la Patria system. Admins in the Tuiteros Activos Telegram group ($\S$~\ref{sec:methods:quantitative}) confirmed that the activity rewards are given on a weekly basis. Figure~\ref{fig:tweet:retweet:stacked:absolute} ($\S$~\ref{sec:results:strategies}) further confirms that several interview participants had significant posting activity levels.

Several participants confirmed reports of the government's use of food distribution as a form of social control~\cite{Renato2020,latimes2019}. Some claimed to receive monthly rewards (payments and food packages) also from individual politicians and organizations, e.g, {\it ``They ask for publicity and offer one bag of food monthly with groceries, vegetables, proteins''} (PG9).

Seven participants claimed to receive no benefits for their activities. Six, both opposition and pro-government, explained that their motivation stems from strong political views and the desire to reveal the real situation in Venezuela, e.g., ``{\it I only do this when I am at mad at the government, as a way to criticize}'' (OP1), {\it ``it is my mission to highlight the advantages of this political, social and economic system for us, the majority, who have been traditionally excluded''} (PG7).

%``{\it if things were good in my country, I would not post anything political. I do politics because political events cost lives at this time in Venezuela''} (OP3),

These findings confirm the classification in $\S$~\ref{sec:results:participants}: participants in Venezuelan campaigns include paid and coerced operatives, and (at least part-time) grassroots campaigners. Our findings are also consistent with recent reports that campaigns are recruiting real people into their operations~\cite{FacebookIO}.

%suggest that participants in Venezuelan campaigns include (1) paid operatives, who receive rewards for social network activities that include distributing disinformation, (2) coerced operatives, and (2) true believers who either do not get paid or are unwilling to be hired for operations that contradict their beliefs.

\subsection{RQ2: International Influence Operations}
\label{sec:results:international}

\noindent
{\bf Past Involvement: Spanish-Speaking Countries}.
Eleven participants, on both sides, claimed to contribute to campaigns for other Spanish-speaking countries. They explained that the contributions included (1) promoting certain hashtags, e.g., ``{\it I worked the coup in Bolivia. We normally have a hashtag, something like \#EvoEsPueblo [Evo Morales]}'' (PG4), (2) tweeting, e.g., ``{\it I publish political tweets for other countries when I see the risk, in this case, I see an extreme risk in Spain with Podemos}'' (OP3), and (3) retweeting, e.g., ``{\it I retweet the Nicaraguans, the Ecuadorians, the Cubans}'' (PG8).

Most of these participants receive requests for help on their communication groups, from operatives in other countries. One participant finds and contributes to campaigns based on interest. None of the participants mentioned receiving explicit benefits for contributions to foreign campaigns. However, for Patria systems users, these activities may count toward their quota for, e.g., activity awards.

\begin{figure}
\centering
\includegraphics[width=\columnwidth]{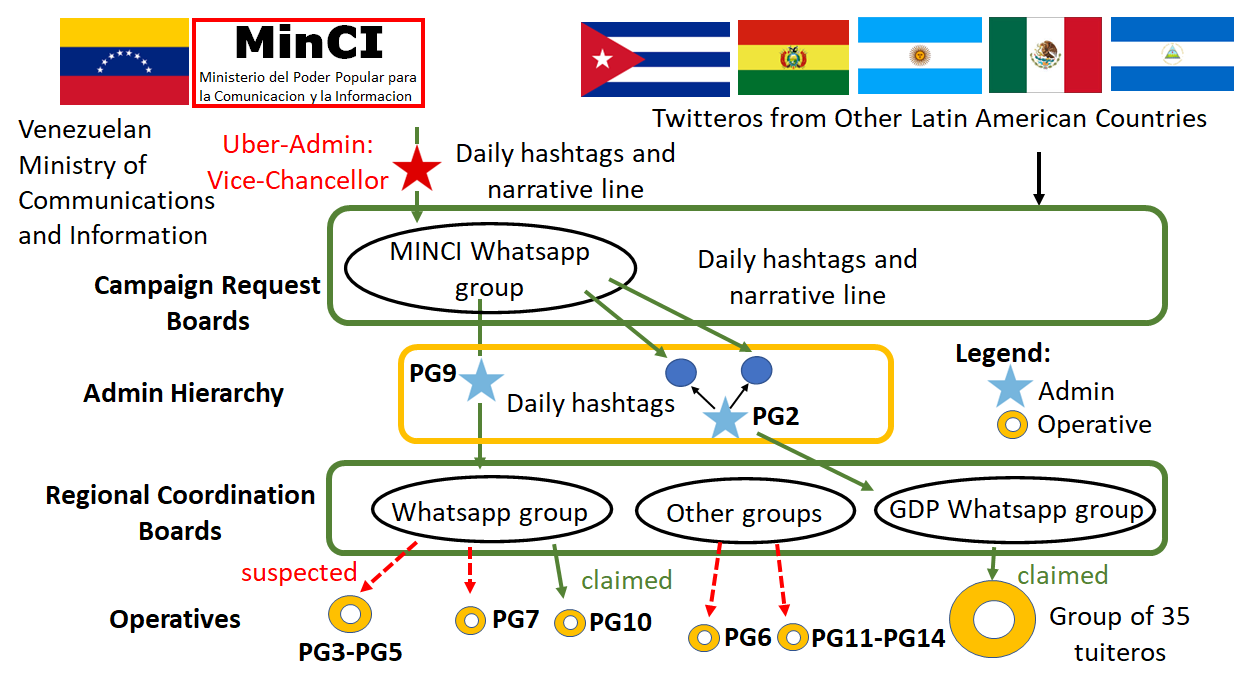}
%\vspace{-20pt}
\caption{Organization structure inferred from pro-government participants. Campaign requests originate from MIPPCI and other Latin American groups, and are communicated and distributed to operatives through online groups organized by a hierarchy of admins.}
%\vspace{-15pt}
\label{fig:organization}
\end{figure}

\noindent
{\bf Willingness to be Hired}.
Fourteen participants said they would be willing to be hired to participate in influence operations on Twitter, including for the US. Their motivation included the impact of Trump's politics and the presence of many Latin Americans in the US. Some agreed conditionally, based on (1) payment, e.g., ``{\it if payment is adequate and I can sustain myself, if I can buy a device able to withstand the work}'' (PG5), (2) the campaign's political orientation e.g., ``{\it If it does not go against my opinion}'' (OP4), and (3) the correctness of the information to be promoted (OP3).

Several participants claimed a keen interest in US politics, and a history of posting content on US politics. Figure~\ref{fig:interests} confirms these claims, showing the number of posts tweeted from the accounts of our participants that mention Trump (1,529 posts), Biden (1,141 posts), or Pelosi (49 posts).

\subsection{RQ1: Organization and Communications}
\label{sec:results:organization}

Several participants revealed their organization structure and communication mechanisms. Figure~\ref{fig:organization} shows information revealed by pro-government participants. Some report and receive instructions from the MIPPCI ($\S$~\ref{sec:methods:quantitative}) through a selective Whatsapp group: {\it ``I am a member of the MIPPCI WhatsApp group. {\bf We use it to plan the hashtags for the next day}. We are a group of around 200 people. Membership is selective, admission decisions are made by the vice-chancellor''} (PG9).

Several participants reported that requests also come from other countries (see Figure~\ref{fig:organization}), e.g., {\it ``I am a member of five international WhatsApp and Telegram groups where we share information that comes from different countries. Sometimes the `tuiteros' from  Nicaragua, Cuba, or Bolivia ask us for help and we support them.''} (PG7). This confirms recent reports from Facebook about the emergence of influence operations that target both domestic and foreign audiences~\cite{FacebookIO}.

Two participants claimed to be {\it admins}, who organize other members through groups that promote each other's political content in Twitter (see regional boards in Figure~\ref{fig:organization}). Consistent with previous findings~\cite{BloombergGuide, OC19}, they revealed hierarchical organizations: {\it ``I am the admin of [anonymized group]. I have 3 sub-admins, and a total of 35 people under my command. [..] {\bf We are organized by regions at the national level.}''} (PG2).

Admins use these groups to (1) distribute the narrative line from the MIPPCI group, e.g., {\it ``I tell my admins what they are going to do, and they are in charge of bringing that message down to the regions through regional groups''} (PG2), (2) coordinate activities, {\it ``We coordinate there and then we publish in Twitter. {\bf We have shifts in which a certain group promotes content}''} (PG2), (3) identify and nudge members that are inactive, e.g., {\it ``I evaluate daily to verify who is working and who is not. If someone is not active, I call them up''} (PG9), and (4) find new clients, {\it ``we talk about content and we even run into clients there''} (PG11).

%Admins explained that some groups have rules, which are communicated to new members, e.g., sending at most three messages per day, and focus on political topics.

Grassroots participants, on both sides, also revealed less structured coordination, where they contribute to the efforts of multiple groups, e.g., {\it ``We started with direct messaging groups from Twitter, we exchanged numbers, and we started creating Whatsapp and Telegram groups, we even have groups in Facebook and Instagram''} (PG5), ``{\it I have three little WhatsApp groups that we created ourselves, where we sometimes share a tweet and we give retweets among all of us}'' (PG4). In particular, opposition participants claimed to post content on their own, and work without an admin: ``{\it {\bf we work alone but together}, we do not have an organization, we do not know each other but we have the same interests''} (OP3).

For both pro-government and opposition reports, we observe consistency with the ``communication constitutes organization'' perspective of organizational theory, that communication and organization co-produce and co-adapt~\cite{PN09}. Similar to volunteer organizations in disaster response~\cite{SP13}, the loose coordination of the opposition enabled them to evolve into an effective organization that distributes information, garners engagement, and promotes hashtags to trending status ($\S$~\ref{sec:results:strategies}).

\begin{figure}
\centering
\includegraphics[width=0.89\columnwidth]{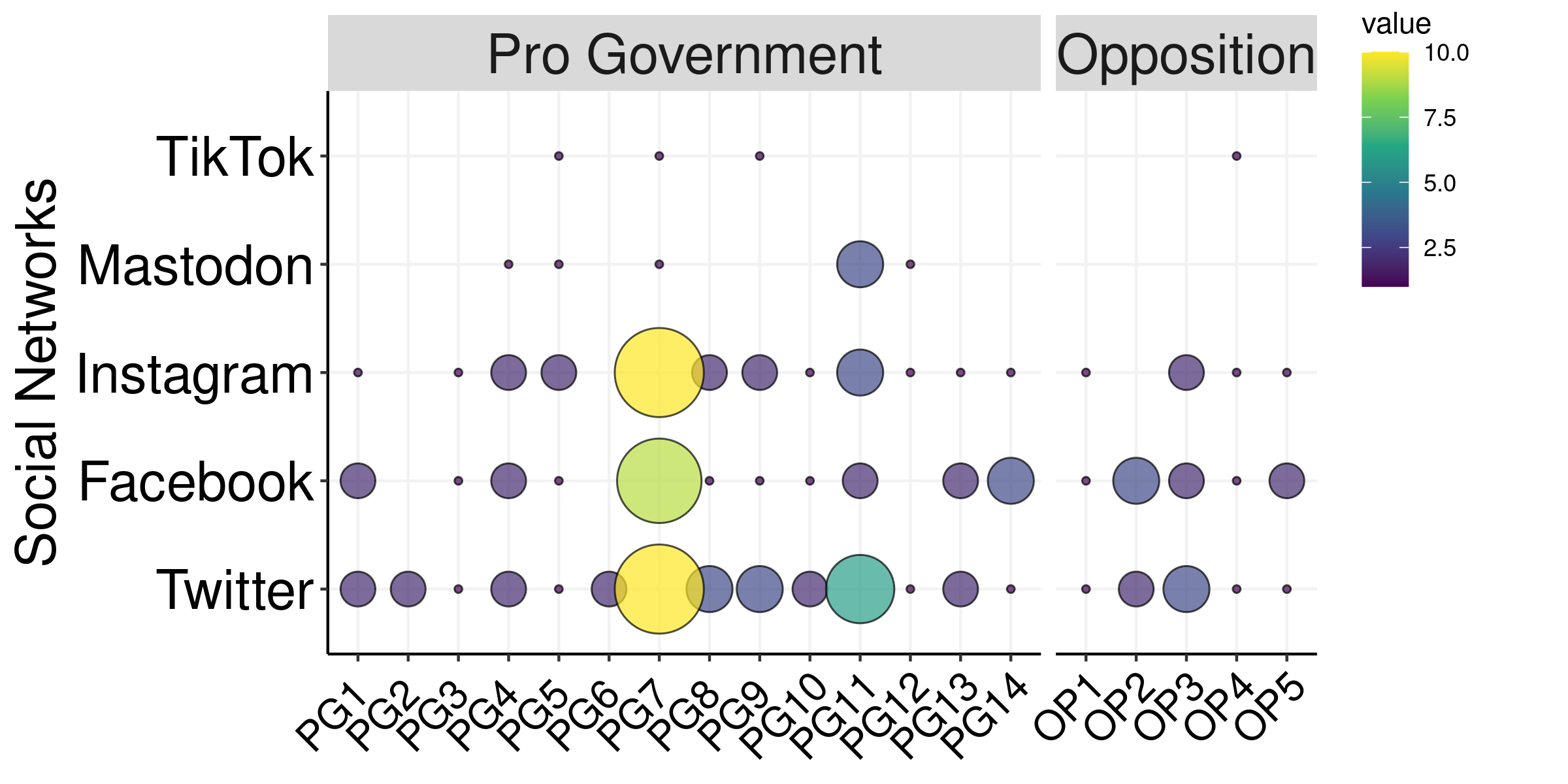}
%\vspace{-10pt}
\caption{Per-participant number of accounts controlled on social networks. Dot sizes are proportional to the number of accounts. A few participants revealed sockpuppet accounts, but most claimed to own only backup accounts.}
\label{fig:accounts}
%\vspace{-15pt}
\end{figure}

These findings confirm that influence operations are collaborative work~\cite{SAW19, FacebookIO, OC19, BloombergGuide}, whether through hierarchical structures consistent with previous reports~\cite{BloombergDocument, OC19}, or through flexible, decentralized structures. The decentralized infrastructure claimed by opposition participants is also consistent with their claims of persecution by the government ($\S$~\ref{sec:results:defenses} and $\S$\ref{sec:results:avoidance}) and the documented news-accessing reliance of the general population on social networks and mobile apps~\cite{NewsBattle, InternetGrip}.

\subsection{RQ1: Capabilities}
\label{sec:results:capabilities}

We discuss participants' insights on accounts and followers.
%
%We discuss their Internet connectivity and device affordability constraints in Appendix~\ref{appendix:findings:limitations}.

\noindent
{\bf Social Network Accounts}.
Figure~\ref{fig:accounts} shows the social networks where our participants claimed to be active, and the number of accounts they claimed to control on each social network. Two participants revealed control of sockpuppet accounts, e.g., ``{\it I have three accounts plus three institutional accounts for which I am the community manager. I also have a personal account.}'' (PG11). PG7 claimed to have 9-10 accounts on each of Twitter, Facebook and Instagram. Three other participants each have three Twitter accounts.

Reasons for having multiple accounts include (1) separating personal from institutional accounts, e.g., ``{\it One is institutional and the other is not so much political but I publish different things}'' (PG13), and (2) separating political from personal accounts, e.g., {\it ``I have family members with different political beliefs [..] {\bf I created separate accounts so to not impose my political messages onto them}''} (PG4).

In contrast, other participants control only one or two accounts in each social networks. Most explained that at most they have a backup account, in case of account suspensions, e.g., ``{\it I have only two, my ``hard'' account, and another account that I have, just in case, because Twitter treats us badly.}'' (PG1). Some explained that this was due to the difficulty of managing multiple accounts: {\it ``I can barely manage two accounts, I cannot imagine how it would be like with many accounts, I wouldn't be able to do it''} (PG1).

Eleven participants revealed control of additional 15 Twitter accounts during the interview. We have manually compared these accounts against their 19 recruitment-validated accounts ($\S$~\ref{sec:methods:quantitative}). We confirmed that with the exception of the two accounts revealed by PG11, all participants use their online identity on the account profile and/or Twitter handle.

\begin{figure}
\centering
\includegraphics[width=0.49\textwidth]{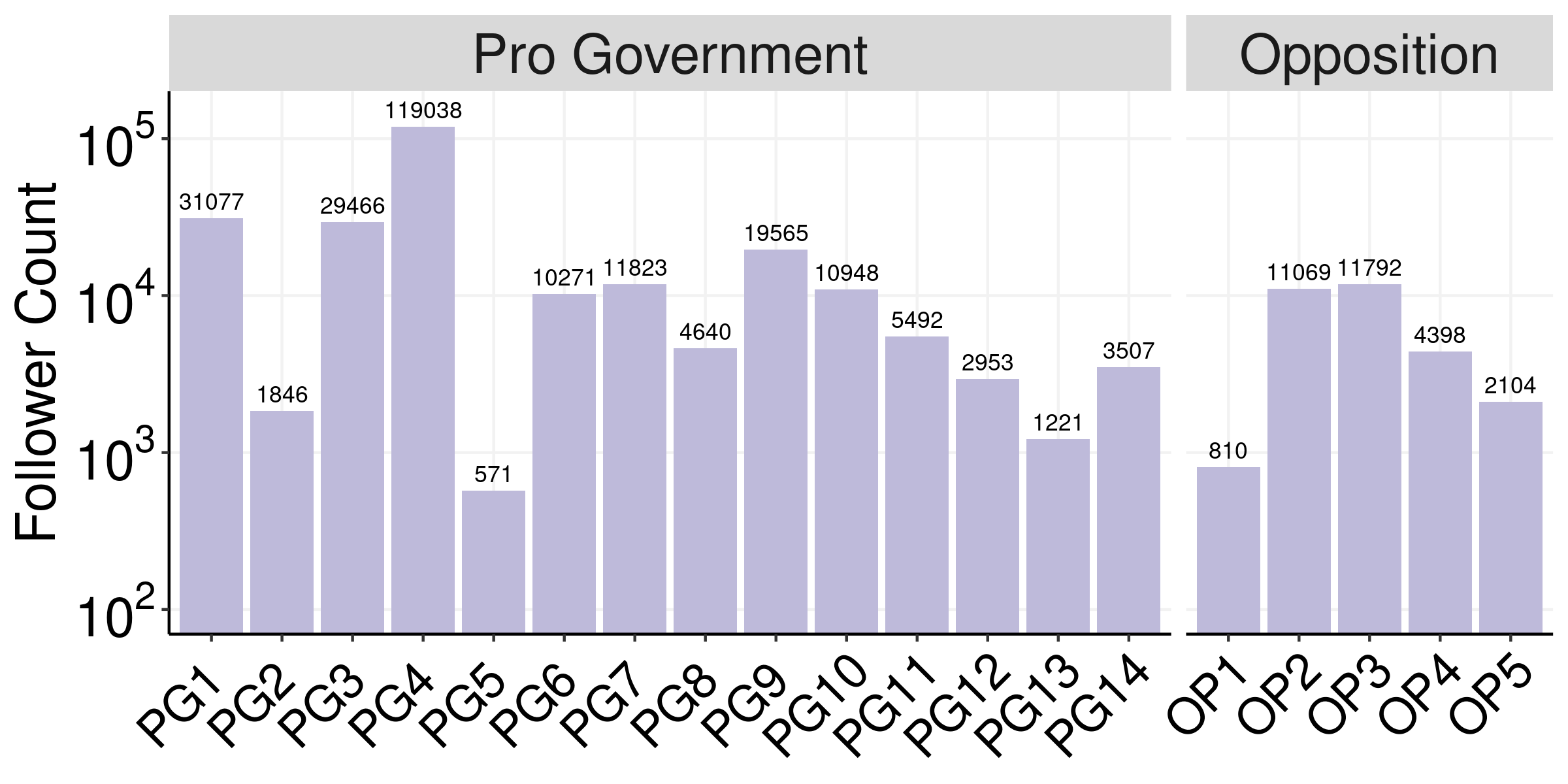}
\caption{
Number of followers for each participant at the beginning of the data collection. The y axis is in log scale. Nine participants have an audience of more than 10,000 followers.}
%\vspace{-15pt}
\label{fig:followers:start}
\end{figure}

This suggests diverse strategies among our participants. While a few rely on sockpuppet accounts, confirming previous findings~\cite{OC19, KCLS17, BloombergGuide}, we found many participants who only control a few personal accounts. This is consistent with participant reports of their use of the Patria system, where they need to register their Twitter account with the platform in order to receive rewards for their activities. This further supports recent Facebook reports that campaigns are starting to recruit real people into their amplification operations~\cite{FacebookIO}.

\begin{figure}
\centering
\includegraphics[width=0.89\columnwidth]{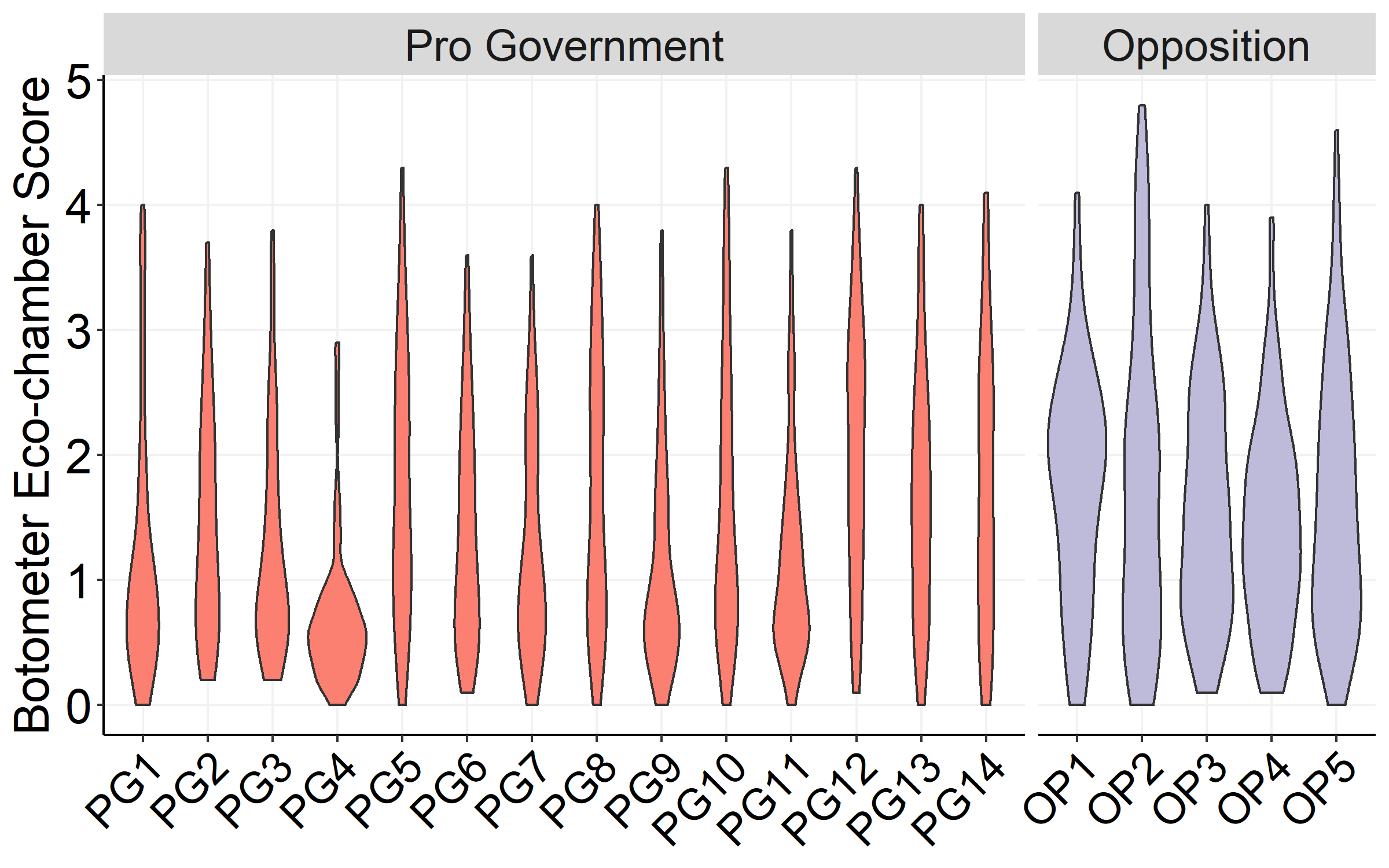}
%\vspace{-5pt}
\caption{Per-participant distribution of Botometer echo chamber scores for a random sample of their followers.}
\label{fig:botometer:echo}
%\vspace{-5pt}
\end{figure}

\noindent
{\bf Followers}.
The followers of an account are vital for its impact.
The numbers of followers revealed by participants are consistent with the ones we collected from their accounts. During several interviews, participants logged into their accounts in order to quote accurate numbers. Figure~\ref{fig:followers:start} shows the number of followers that we collected on February 2nd, 2021 from each participant. Our participants can reach a large audience: Nine participants had more than 10,000 followers, with the maximum being 119,038 followers (PG4). 

To evaluate the ability of our participants to reach a wide audience, we used the Botometer tool~\cite{SVYFM20} on a random sample of 100 followers from each participant. Botometer provides scores on a 0 - 5 scale, where high scores denote more likely bots, and scores in the middle denote uncertainty~\cite{Botometer}.

\begin{figure}
\centering
\includegraphics[width=0.89\columnwidth]{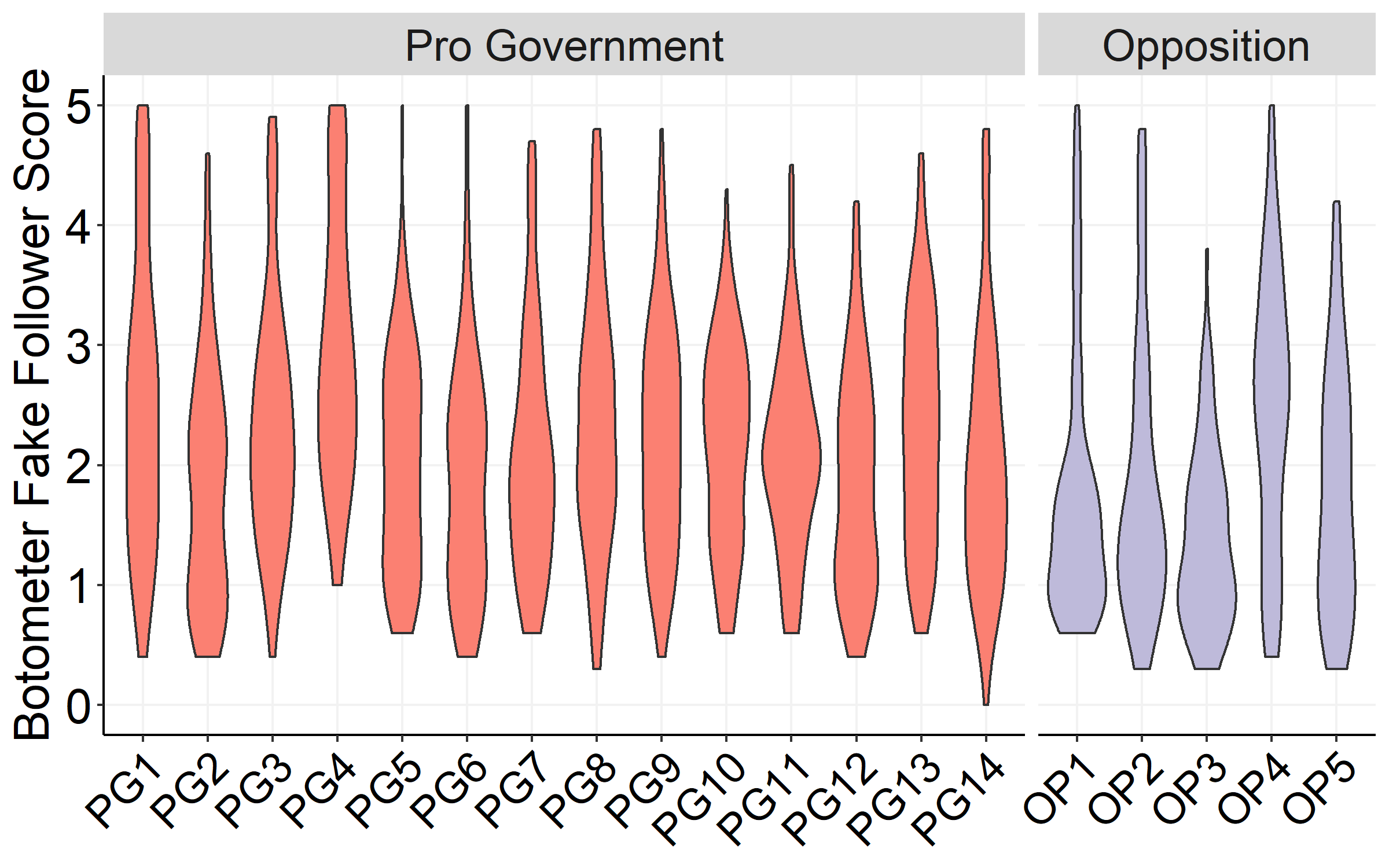}
%\vspace{-5pt}
\caption{Per-participant distribution of Botometer fake follower scores for a random sample of their followers.}
\label{fig:botometer:fakes}
\end{figure}

Figure~\ref{fig:botometer:echo} shows the per-participant distribution of their followers' Botometer echo-chamber scores (0 - 5 scale). This score signals accounts that engage in follow-back groups and share and delete political content in high volume~\cite{Botometer}. Between 0 and 22.93\% of the participants' followers had scores of at least 3 (M = 9.47, SD = 6.01). Figure~\ref{fig:botometer:fakes} shows the per-participant distribution of Botometer fake follower scores for the same random sample of their followers. These scores identify bots purchased to increase follower counts~\cite{Botometer}. Between 2\% and 43.62\% of the participants' followers had scores of at least 3 (M = 19.19, SD = 9.49).

Participants had 56.38\% to 92\% (M = 72.32, SD = 9.65) followers with both scores under 3. This suggests that while some participants had many suspicious followers, most also have significant numbers of genuine followers, and may reach a wide audience.

Several participants explained that their follower communities are smaller than what they should be, due to Twitter (1) removing subsets of them, (2) suspending their accounts, or (3) due to periods of inactivity, e.g., {\it ``I was outside of Twitter for a year, because I was a political prisoner. During that time, many people stopped following me.''} (OP2)

\subsection{RQ1: Political Promotion Strategies}
\label{sec:results:strategies}

We discuss strategies to create and promote political content.

%We include more strategies in Appendix~\ref{appendix:findings:promotion}.

\noindent
{\bf Daily Hashtags: Creation and Promotion}.
An admin participant provided insights into the creation of the daily hashtags, by the MIPPCI board where he is a member: {\it ``We make hashtag proposals daily based on the political movement of the day. Once every hashtag has been proposed we start studying them [..] and cast our votes until we reach a consensus. {\bf The vice-chancellor has the last word}''} (PG9).

Admins distribute these hashtags through their groups, see Figure~\ref{fig:organization}. Several participants explained that they monitor the posting of these daily hashtags, to include in their tweets in order to simultaneously (1) promote the hashtags, e.g., {\it ``Everything I publish I accompany with the hashtag from the MIPPCI. We use the hashtag so that it gets more interaction''} (PG10), and (2) garner engagement for their own posts, {\it ``If you take advantage of those first 5-10 mins after the hashtag is announced, the post will receive support [engagement] throughout the day. You can get up to 700 retweets''} (PG10).

We believe that the goal of these efforts is not only to make hashtags reach trending status, but also to maximize the time they stay in the top trending list: an account that uses a hashtag after it reaches a high rank, helps the hashtag stay trending.

%Figure~\ref{fig:trolls} in Appendix~\ref{appendix:findings:promotion} illustrates the hashtag promotion protocol for pro-government accounts, with snapshots that we took on the ``Tuiteros Activos'' Telegram group. We observe the speed of bringing the promoted hashtag to trending status.

\begin{figure}
\centering
\begin{minipage}[b]{0.39\textwidth}
\includegraphics[width=\columnwidth]{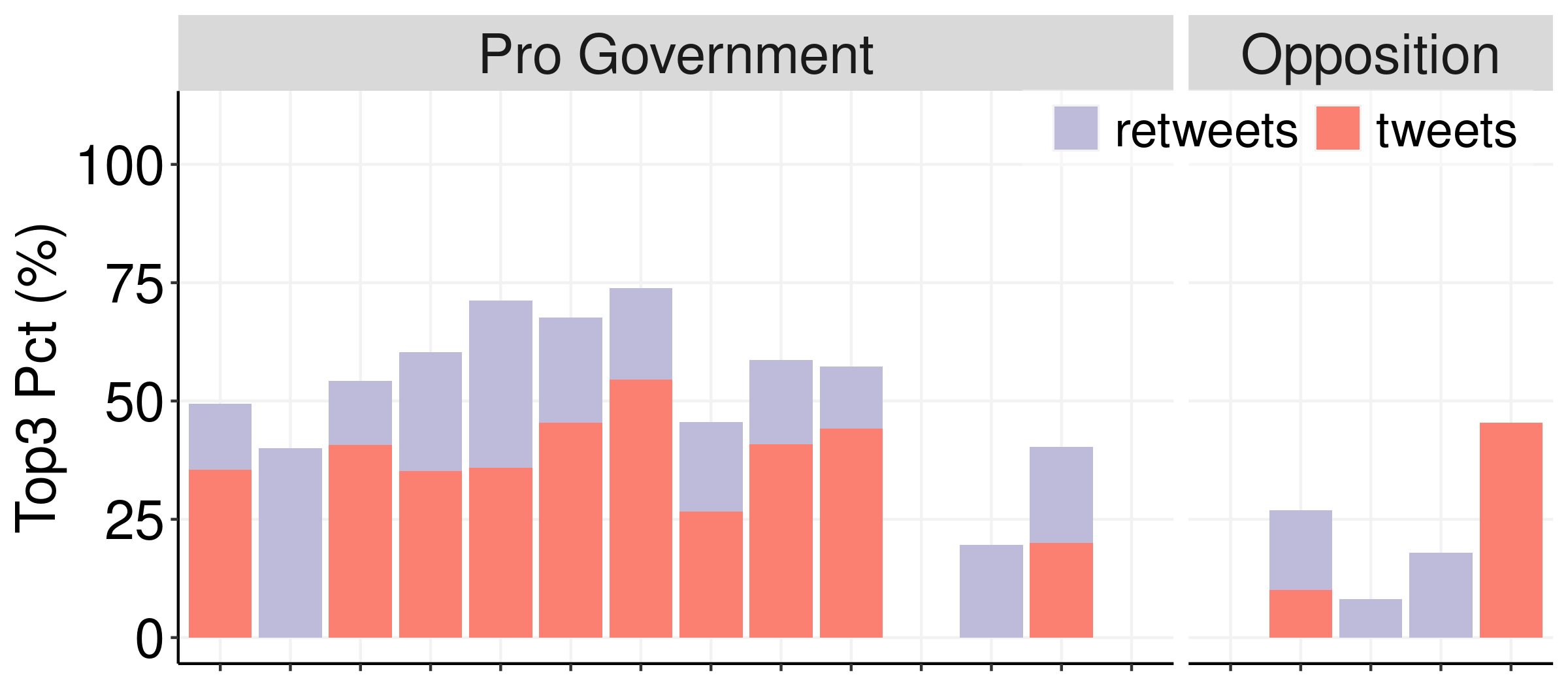}
\includegraphics[width=\columnwidth]{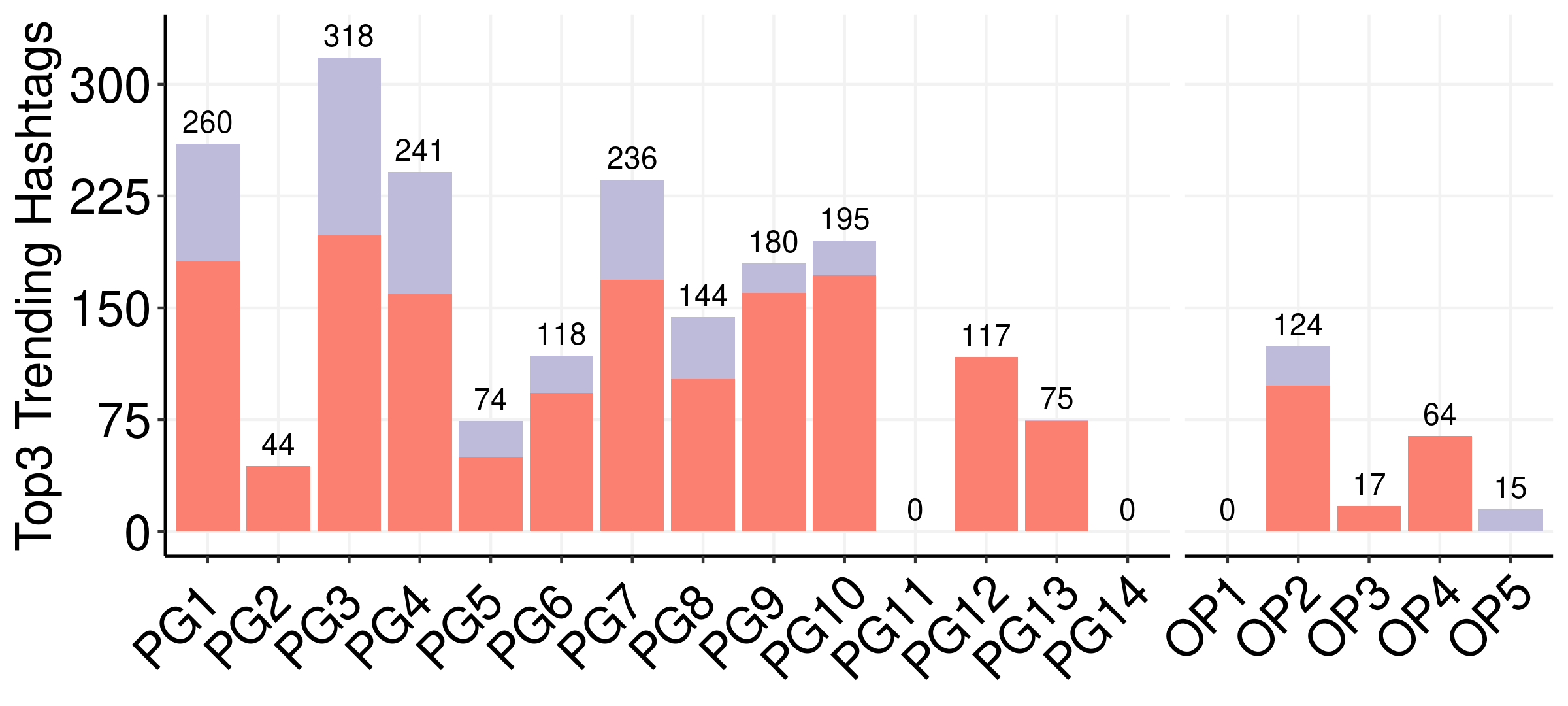}
\end{minipage}
%\vspace{-10pt}
\caption{Trending hashtags. Top: Per-participant percentage of hashtags that appeared in a post, and were top-3 trending anywhere in Venezuela. Bottom: Absolute hashtag counts.
%
%Eleven participants included top-3 hashtags in their original tweets; 12 participants retweeted posts that already had these hashtags.
}
%\vspace{-5pt}
\label{fig:hashtag:trending}
\end{figure}

We plotted the percentages and absolute counts of hashtags that (1) appeared in participant tweets or retweets in a three month interval, and (2) have become top-3 trending hashtags anywhere in Venezuela. Figure~\ref{fig:hashtag:trending} shows that 16 of the 19 participant-controlled accounts, posted either a tweet or a retweet containing a hashtag that reached top-3 trending. Eleven participants posted at least one original tweet with a top-3 hashtag; up to 55\% of hashtags included by one participant (PG7) in original tweets had a top-3 trending rank.

These hashtag-promoting attacks compromise information integrity because they fraudulently promote the search rank of desired hashtags, and simultaneously demote other hashtags promoted (perhaps organically) by opposing campaigns. We note that this attack differs from the ephemeral astroturfing attacks of Elmas et al.~\cite{EOOA21} in that it (1) involves humans instead of astrobots, and (2) does not seek to remain invisible, e.g., by erasing hashtag-promoting tweets.

%\nestor{good opportunity to cite ~\cite{EOOA21} where they study artificial hashtag promotion on Twitter, found that at least 20\% of the top 10 global trends are subject to this type of attack. }

\noindent
{\bf Content Creation vs. Engagement: Perception}.
Sources of inspiration for the content created for original tweets include the mission and vision of the client (for participants who claimed to work for various clients) and also news portals:
%
%\blockquote{
{\it ``We access news portals like RT Actualidad [Russian news outlet], HispanTV [Iranian news outlet in Spanish], and CNN''} (PG10). We confirm that 219 posts of our participants included links to Russian~\cite{ActualidadRT, SputnikMundo}, Iranian~\cite{HispanTV} and Cuban~\cite{CubaInformacion, CubaSi} news outlets, and 559 posts had links to government sites. These behaviors confirm strategies of integration of government and externally-funded media as source content for narratives, previously reported in countries like Syria~\cite{SAWVYS18, SAW19}.

Pro-government participants confirmed their use of information distributed through MIPPCI, e.g., {\it ``We receive daily the information from the MIPPCI. They say, look, today’s line is this .. and so we read the news release and we compose the final content with our own words.''} (PG2).

Participants explained the importance of the audience reached and engagement received (number of times people interacted with their tweets, i.e., retweets, quotes, replies and likes) by their tweets. They are used by admins for evaluation purposes, e.g., ``{\it admins look at the amount of retweets that I received, the amount of followers that the account has gained, the projection of the publications''} (PG11), and also by the Patria system~\cite{Patria} ($\S$~\ref{sec:results:motivation}) to assign bonuses, e.g., ``{\it My account is mentioned quite a lot, and the more engagement I have, this bonus arrives without me needing to publicize''} (PG10). 

\begin{figure}
\centering
\includegraphics[width=0.45\textwidth]{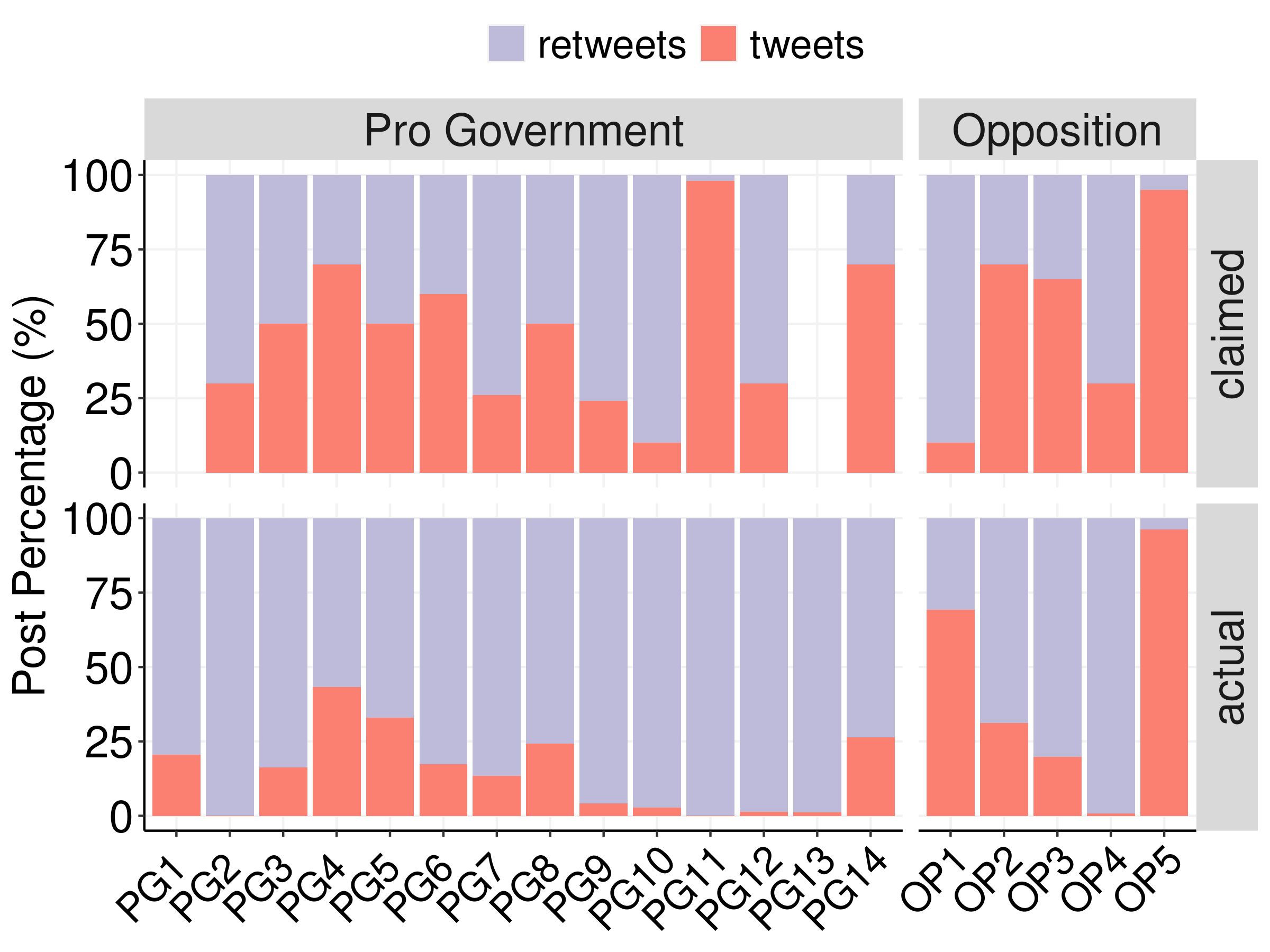}
%\vspace{-5pt}
\caption{Participant tweets (red, bottom) vs. retweets (purple, top). Participant-claimed percentages top, real percentages bottom. PG1 and PG13 did not provide an answer.}
\label{fig:tweet:retweet:stacked:percentage}
%\vspace{-5pt}
\end{figure}

A widespread theme is a peer-based strategy to acquire engagement, where participants share their tweets in communication groups, and retweet the tweets posted there by other members, e.g., ``{\it When I see a tweet from someone in my group, I retweet immediately. This is the work each and every one of us does.}'' (PG2), ``{\it I am a member of six groups, I go into each group and I retweet whatever they publish during the day, no matter the content.}'' (PG10). Participants explained that they expect to receive engagement for their tweets, once they share them in their groups. This likely results in lockstep behaviors, which can be exploited to detect influence operations ($\S$~\ref{sec:assurance}).

Participants further reported unexpected strategies, i.e., (1) receiving requests for retweets through direct messages, (2) retweeting their own tweets, and (3) mentioning select accounts in their tweets to encourage reciprocation:
%
%\blockquote{
{\it ``On every tweeted news I mention at least six accounts of people that follow me and consistently retweet the information that I post. I look up the number of their followers so that I know that it is worth mentioning their account
%
%If previously my news reached 1,000 people directly, [with this technique] in the course of an hour it will reach indirectly around 2000 or 3000 people more.
%
''} (OP2).

We confirmed that most participants have posted or retweeted such content. Some explained that they preferentially retweet posts from certain sources, e.g., {\it ``We retweet the information from the government work and political figures, and show the work done by state institutions''} (PG11). This explains our reports of hyperpartisan news with images from staged events ($\S$~\ref{sec:results:fakenews}). The ``Tuiteros Activos'' Telegram group ($\S$~\ref{sec:methods:quantitative}) had claims that the Patria system gives bonuses only for retweets of accounts controlled by government officials.

We further investigated the participant perception of the distribution of their original posts versus retweets. Figure~\ref{fig:tweet:retweet:stacked:percentage} (top) shows claimed percentages. A majority of participants claimed to post a mix of original content and retweets, thus to be both content and engagement creators.

One participant motivated this strategy by the need to appear influential to clients, {\it ``It does not look good if the people that hire us see that all we do is retweet. So, we try to have more tweets than retweets in our main accounts''} (PG9). Two opposition participants said they post more retweets due to self-censorship, e.g., {\it ``I sometimes express an opinion, but very little, because they punish people, politically. If you go directly against the government, then they look for you''} (OP4).

\begin{figure}
\centering
\includegraphics[width=\columnwidth]{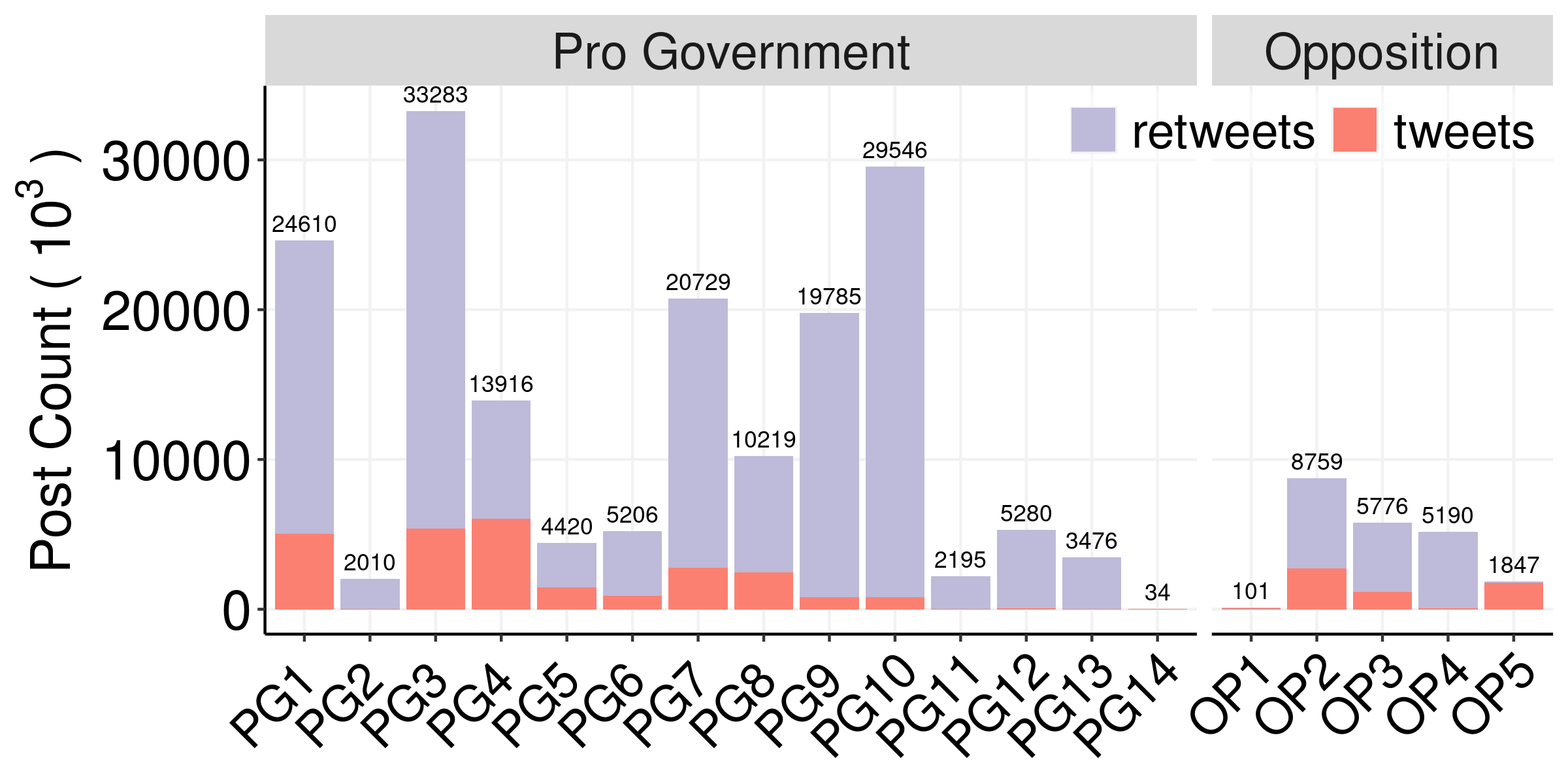}
%\vspace{-20pt}
\caption{Number of tweets vs. retweets collected from participant accounts over three months. We observe substantial efforts to create engagement for content posted by others.}
%\vspace{-5pt}
\label{fig:tweet:retweet:stacked:absolute}
\end{figure}

\noindent
{\bf Content Creation vs. Engagement: Twitter Truth}.
To take steps toward verifying several of these claims, we collected all the Twitter posts of the interview participants over three months. Figure~\ref{fig:tweet:retweet:stacked:percentage} (bottom) shows the real percentages during this interval. Figure~\ref{fig:tweet:retweet:stacked:absolute} shows the absolute values. Only two participants (OP1 and OP5) posted more original tweets than retweets. In fact, seven participants have posted less than 65 original tweets each, over three months.

\begin{figure}
\centering
\includegraphics[width=\columnwidth]{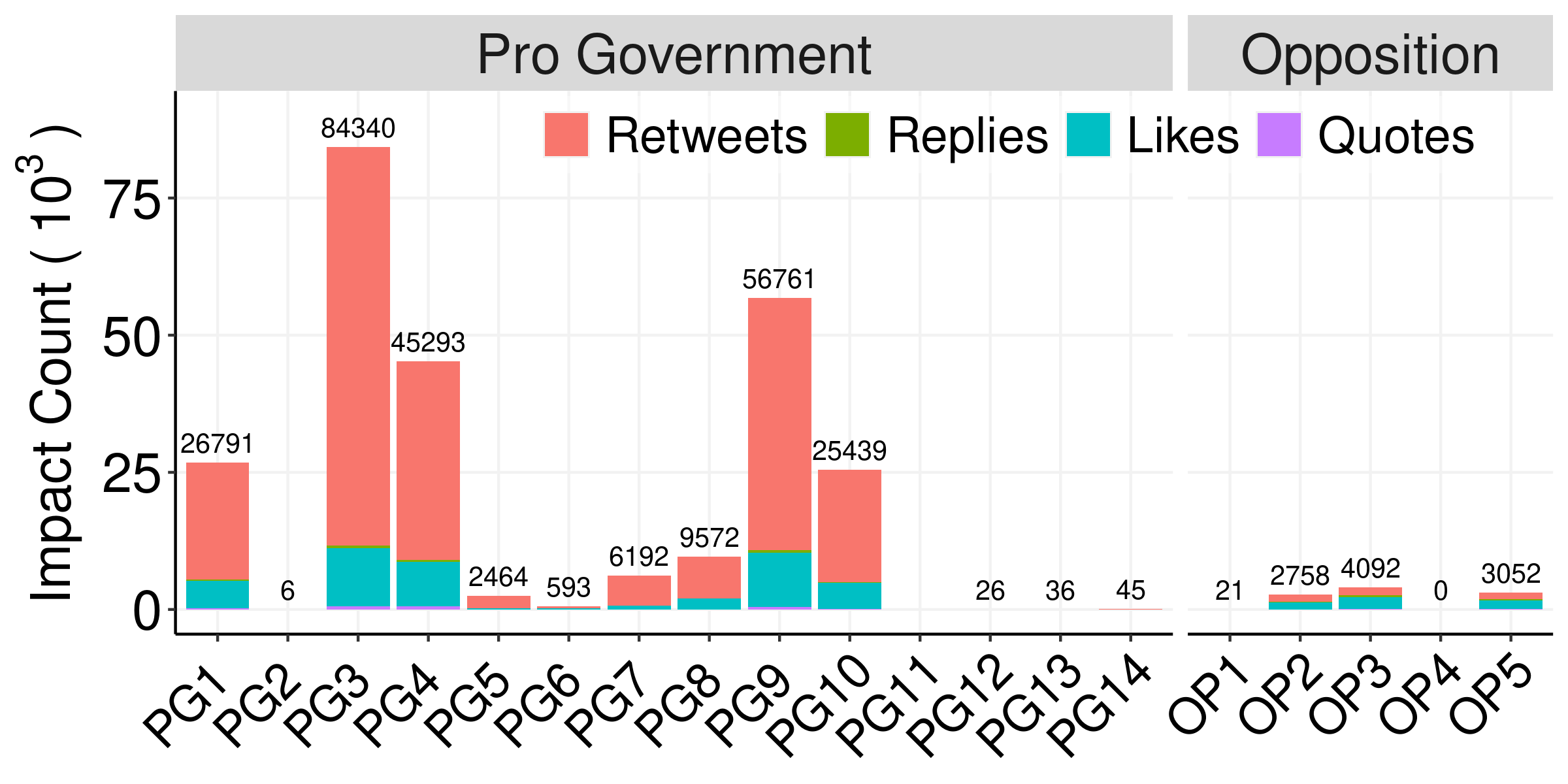}
%\vspace{-25pt}
\caption{Engagement received over one month, by original posts of our participants. We observe influencer potential or a well-oiled propaganda machine, for several participants.}
%\vspace{-10pt}
\label{fig:tweet:impact}
\end{figure}

To analyze engagement claims, we collected the per-participant engagement, i.e., the total number of retweets, replies, likes and quotes received by the original tweets posted from their Twitter accounts during a one month interval. We collected this engagement almost one month after the end of the posting interval. Figure~\ref{fig:tweet:impact} plots this data. PG11's account was suspended on the day we collected this snapshot and OP4 had not received any engagement during the period examined. We observe that the posts of five participants received towering engagement, each with a total over 25,000. PG3 had a one-month engagement of 84,340. The average per-tweet engagement of PG9 and PG10 was 291 and 189 respectively.

\subsection{RQ4: Exploration of Twitter Defenses}
\label{sec:results:defenses}

Interview participants reported a suite of penalties they experienced in Twitter. We discuss these in the following.

\noindent
{\bf Account Closure, Restriction, Suspension}.
Most participants confirmed to have had at least one account suspended by Twitter.
%
%A few claimed that this is infrequent, e.g., ``{\it I have been restricted [in Twitter], maybe a couple of times. They send me the code and everything is solved}'' (PG3).
%
Several reported frequent suspensions, that prevent them from accessing their accounts for days, e.g., {\it ``That account got blocked, suspended, restricted, it used to be between 5 days to a week when I could not use it''} (PG4). During our monitoring interval, we recorded five suspension events for the accounts of our participants.

Several participants revealed that Twitter directly closed their accounts, e.g., {\it ``I had another account that got canceled, not even suspended. The SEBIN [Venezuela's political police] objected [to Twitter] and asked information about the account. Twitter didn’t give any information to SEBIN. Thank God, otherwise I wouldn’t be talking with you today''} (OP3).

\begin{figure}
\centering
\includegraphics[width=0.47\textwidth]{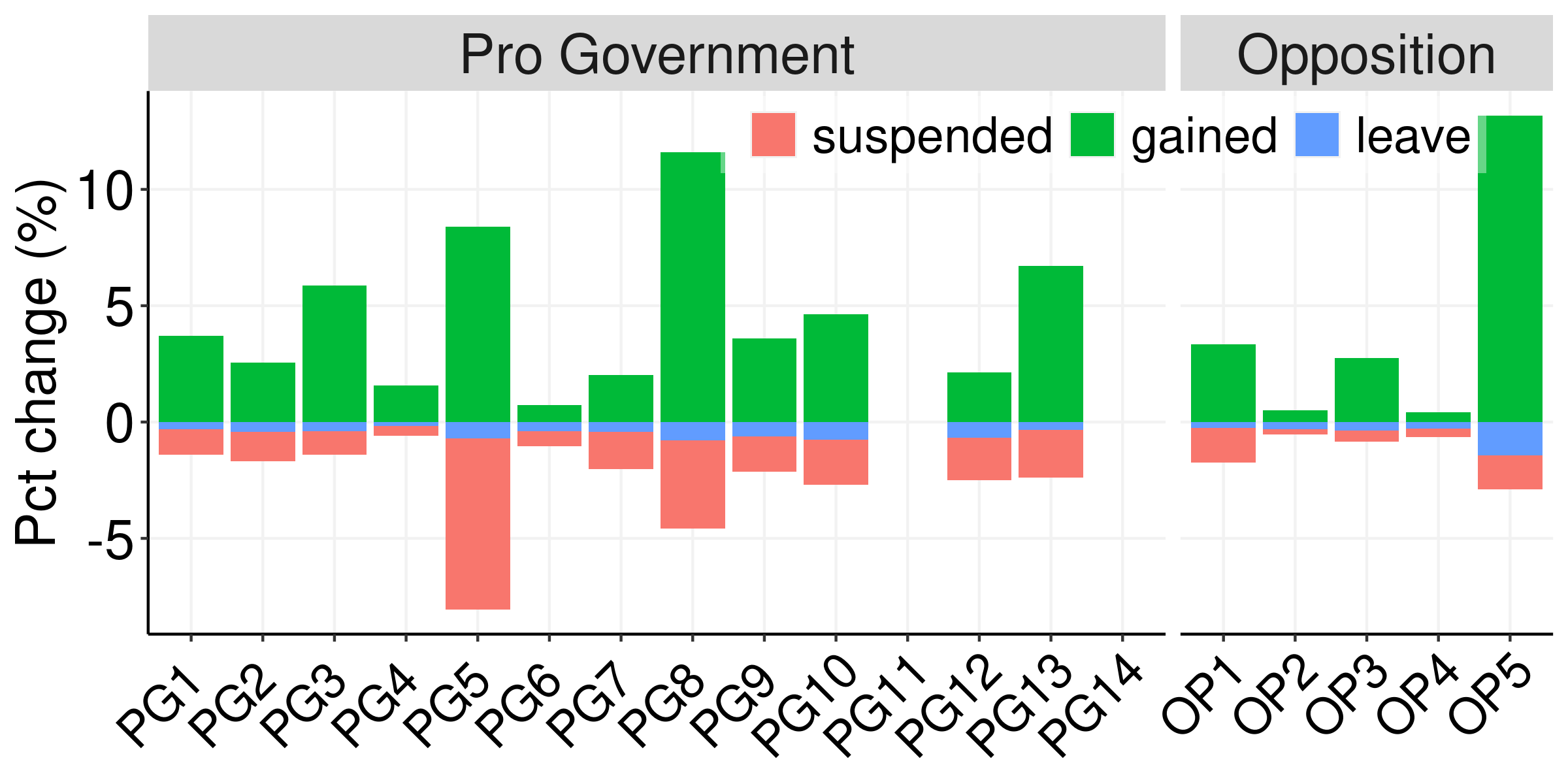}
\vspace{-10pt}
\caption{Follower percentage change from the initial count, after two months. Overall, most participants gained followers.}
\vspace{-15pt}
\label{fig:followers:changes}
\end{figure}

\noindent
{\bf Pruning of Followers}.
Two participants claimed that Twitter removed followers from their accounts on multiple occasions, e.g., {\it ``My account used to have 8,500 followers and after the first suspension it had 1,100 followers. After that, it reached 12,000 and after the second suspension they returned it with 9,800.''} (PG10). He surmised that such events could also occur because some of his follower accounts were suspended by Twitter. Figure~\ref{fig:followers:changes} shows the per participant number of follower accounts that were suspended by Twitter during a two months interval. It provides evidence toward confirming these claims.

\noindent
{\bf Shadowbans and Content Flags}.
Two participants reported that Twitter shadowbanned their accounts. They mentioned Shadowban~\cite{Shadowban}, a webservice popular in their community, to detect if an account has been shadowbanned. One participant described an experiment performed by her Twitter group to discover inconsistent counting of retweets by Twitter: {\it ``We have a group of 50 people. We each posted exactly the same tweet. We then all retweeted that same tweet for everyone. So, every tweet should have 50 retweets. One didn’t reach 10, some reached 20, some other reached 30-40ish''} (OP3).

\begin{figure}
\centering
\includegraphics[width=0.89\columnwidth]{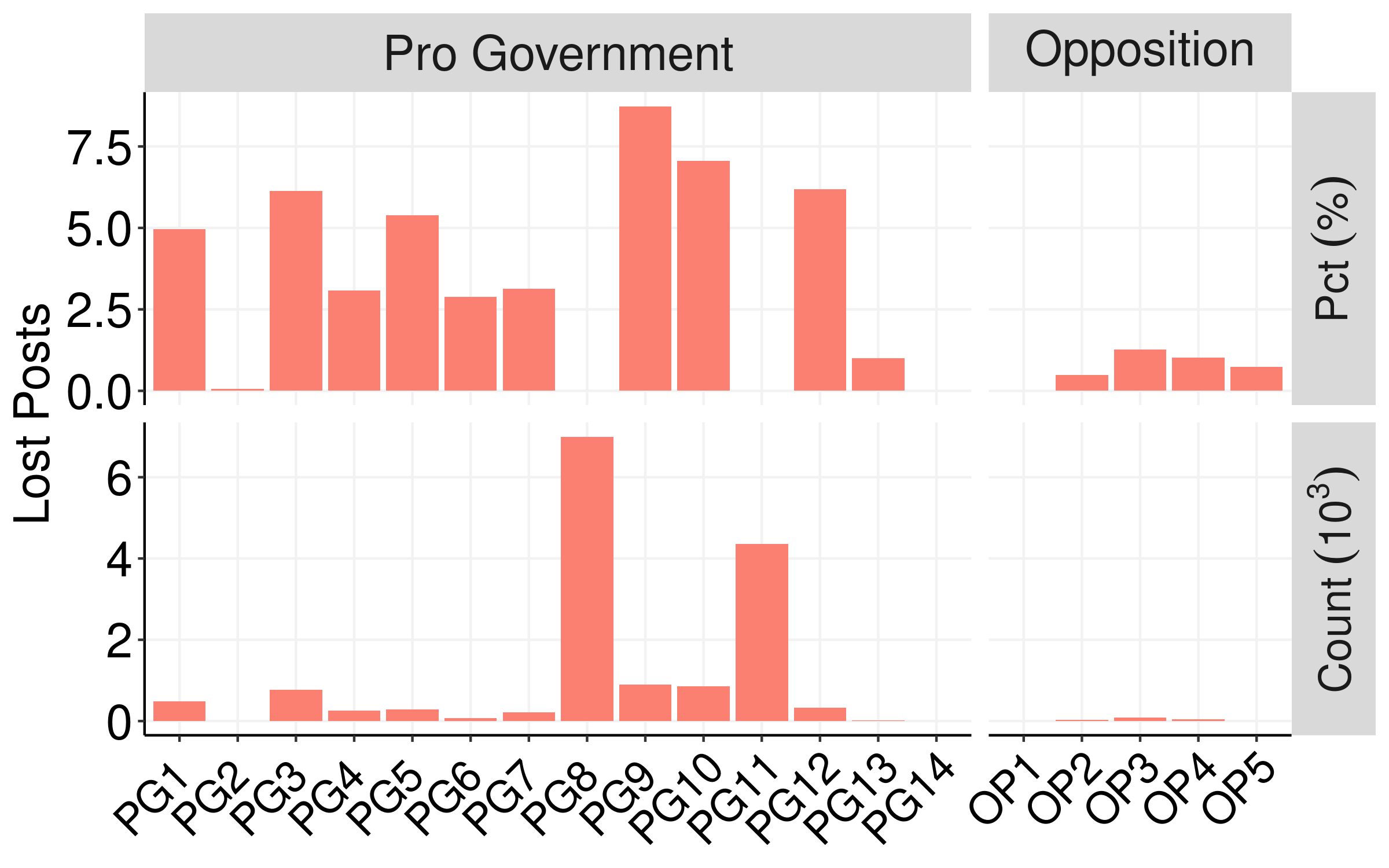}
\vspace{-15pt}
\caption{Number of posts over two months, that were suspended or deleted by Twitter one month later: (bottom) actual values in thousands, (top) percentage from their total posts.}
\vspace{-10pt}
\label{fig:tweets:deleted}
\end{figure}

Figure~\ref{fig:tweets:deleted} shows the number of tweets and retweets that were posted by the participants over two months, and were suspended or deleted by Twitter one month later. Only retweets were removed, and most because the posting account was suspended by Twitter. The accounts of PG8 and PG11 were in a suspended state during this interval.

\subsection{RQ4: Detection Avoidance and Recovery}
\label{sec:results:avoidance}

Participants revealed several strategies to bypass Twitter's penalties. We discuss these in the following.

\noindent
{\bf Rate Limiting Efforts vs. Activity Quotas}.
Several participants explained that the above penalties are due to high posting rates, e.g., ``{\it if I do too many retweets and replies Twitter interprets this as if I was a robot}'' (PG5). Even short bursts of tweets can lead to account suspensions, ``{\it after the 6th or 7th tweet, bam, my account would get restricted.}'' (PG4).

To avoid detection, participants report limiting their posting rates, e.g., {\it ``When I get to around 20 retweets I stop, I logout of the account because the limitation may be about to pop up''} (PG5). They also claimed to space out their posts, e.g., {\it ``In the morning I posted 20 tweets. Then I wait 2 hours, and I post 20 more''} (PG11). This is consistent with strategies shared on Telegram groups, of waiting at least 5s between posts, or posting five tweets slowly every 10 mins.

We find a conflict for participants who reported quotas on their daily original tweets. Quotas include (1) upper bounds, e.g., ``{\it We do not go beyond 10 posts a day when we are hired''} (PG11), (2) lower bounds, e.g., {\it ``Our job is to post 10 tweets a day, but if you want more, it's fine''} (PG2), and (3) contract-based, e.g., {\it ``Depends on the contract I have with my customer, the payment they offer''} (PG9). We exploit this conflict in $\S$~\ref{sec:assurance}.

Figure~\ref{fig:tweet:retweet:stacked:absolute} shows that 17 participants posted more than 10 tweets and retweets on average per day over three months, with nine posting more than 60; PG3 posted 341 daily posts.

\noindent
{\bf Backup Accounts}
Several participants claimed that to recover from longer or permanent suspensions, they create {\it backup accounts} whose names are a variation of their main account name. We analyzed the 15 additional Twitter account handles revealed by 11 participants during interview. These participants revealed each between 1 and 3 additional Twitter accounts. For each of the nine participants whose revealed accounts were accessible, we confirmed that these accounts have either similar Twitter handles or the same screen name. We conjecture that this occurs because of the participant need to confirm identity for the government Patria app and receive bonuses for their online activities (see $\S$~\ref{sec:results:motivation}).

\noindent
{\bf Avoiding Suspension Wars}.
Several participants claimed that Twitter penalties ($\S$~\ref{sec:background:goals}) are due to reports from other accounts, resulting in a {\it suspension war}: {\it ``People can create up to 10 accounts to report someone else. {\bf The government collectives use large networks to annihilate opposition accounts}''} (OP3). Twitter does provide mechanisms for users to report tweets that they consider abusive or harmful~\cite{TwitterReport}. Participants avoid their tweets being reported by others~\cite{TwitterReport} by being nice to their social entourage, e.g., {\it ``I have never blocked or reported anyone. This is my policy. I accept anything, anyone can comment whatever they want. If I do not like what you publish, I do not follow you''} (PG9).

\noindent
{\bf Avoiding Post Automation}.
While a few participants know others who use Hootsuite~\cite{Hootsuite} and Tweetdeck~\cite{Tweetdeck} to schedule tweets, most use only the official Twitter app or Web UI to post content. This is due to fear of detection: {\it ``Twitter closes your account because you robotized. You can detect bots because they are programmed to post at certain times''} (OP2). Our qualitative analysis mostly confirms these claims. Only a few of our participants have used tools to post their tweets: dlvr.it~\cite{dlvr.it} (PG6, 181 tweets), Instagram (PG5, 14 tweets) and Tweepsmap~\cite{Tweepsmap} (PG9, 1 tweet; PG7, 2 tweets). In $\S$~\ref{sec:assurance} we discuss the potential to identify suspected influence operations participants, among accounts that use automation.

Further detection avoidance and recovery strategies include (1) careful management of accounts, IP addresses and browsers, (2) using self-censorship to avoid toxic language and offensive images, and (3) appealing account suspensions.

\section{Information Assurance}
\label{sec:assurance}

We now discuss discovered vulnerability points (VP) of IO participants, summarized in Figure~\ref{fig:strategies}. We then suggest changes to social networks' handling of influence operations.

%We interviewed participants who are flexible and thrive, acquire tens of thousands of followers and generate hundreds of daily posts that receive tens of thousands of retweets and likes.

\subsection{IO Vulnerability Points (VPs)}
\label{sec:assurance:vp}

\noindent
{\bf VP1: Identify and Penalize Daily Hashtags}.
Hashtags promoted by operatives can be found by monitoring communications groups used by operatives ($\S$~\ref{sec:methods:qualitative}). Such groups often have open membership, e.g., ``{\it Maybe I put up a good word about you [the interviewer] with the administrator of the group. Everyone is looking for people that do retweets''} (PG5). Hashtags posted in these groups are available to all members, and can be validated against posts from hyperpartisan Twitter accounts (e.g., @mippcivzla).
%
%The analysis in $\S$~\ref{sec:results:strategies} further revealed that participants often include these ``daily hashtags'' into their tweets and retweets.

\noindent
{\bf VP2: Patria System}.
The Patria system~\cite{Patria} reported by pro-government participants ($\S$~\ref{sec:results:motivation}) uses the Twitter API to link Twitter accounts and to keep track of tweet counts to rank and pay participants. While Twitter could block Patria's access to the Twitter API, a smarter strategy is to determine if apps like the vePatria, veMonedero and/or veQR, are co-installed on the user's device. Twitter can then use other device information (e.g., IP address, model) to identify the Twitter accounts registered on the device.
%
%We note that all our participants claimed to access Twitter from Android devices.
%
Platforms like Twitter can also generalize this approach to identify other apps that use their APIs and are co-installed with their client. This would enable them to detect operatives active through other frameworks and countries.

\noindent
{\bf VP3: Lockstep Behaviors}.
Participants revealed lockstep behaviors that arise from their rush to include newly released daily hashtags into their posts, and their peer-based strategy to acquire engagement ($\S$~\ref{sec:results:strategies}). We observe the opportunity to leverage existing lockstep behavior detection solutions~\cite{CYYP14, SMJEKV15, SLK15, LFWMS17, YMGYLSWL19, HRRC18}, to detect groups of social network accounts with synchronized behaviors.

\noindent
{\bf VP4: Disinformation and Political Outlets}.
Our finding that participants contribute to the distribution of disinformation and articles from hyperpartisan outlets ($\S$~\ref{sec:results:fakenews}) suggests the opportunity to leverage existing efforts to detect misinformation and disinformation~\cite{SHSL21, SWL19, LBBBGMMNPR18},
%
%~\cite{ZCLSZS21, SHSL21, SWL19, RCMVB19, SSWTL17, LBBBGMMNPR18},
%
and identify accounts responsible for posting or promoting such content.

\noindent
{\bf VP5: Activity Quotas vs. Rate Limits}.
In $\S$~\ref{sec:results:avoidance} we found that IO participants need to post and amplify many messages daily, while simultaneously avoiding detection and account suspensions. Even though participants reported limiting their posting rates, our analysis of their accounts revealed that in reality, many continue to post significant daily content over long periods of time. This suggests the ability to detect accounts with suspicious levels of activity, including substantial activity bursts~\cite{RHRAC19, NAGKV19, LFWMS17, HSBGAKMF16, XZLW16, LNJLL10, KCS18, LCNK17}, e.g., associated with the release of the daily hashtags.

\noindent
{\bf VP6: Post Automation}.
To sustain high levels of activity, operatives may use post automation tools, see $\S$~\ref{sec:results:strategies}.
%
%From the 2,918,175 most recent posts of the 1,032 hyperactive followers of @mippcivzla, 99,335 were posted using 31 automation tools that include TweetDeck, IFTTT, Hootsuite, PostCron, Postoplan, Blog2Social, Envio de Tweets. 38 of these accounts have used such tools to post between 1,000 and 3,025 of their 3,025 most recent tweets. These are all established accounts, between 1,773 and 5,067 days old.
%
Participants observed that accounts that use post automation tend to post at predictable times ($\S$~\ref{sec:results:avoidance}). This suggests further opportunities to identify automated accounts.

\subsection{Proposed Next Steps}
\label{sec:assurance:defenses}

Our study reveals that influence operations continue to thrive despite social network defenses that include suspending accounts or shadowbanning posts. Based on our findings, we suggest that social networks could instead implement the following, more nuanced approach toward influence operations, that leverages knowledge of the different IO participant types.

\noindent
{\bf Classify IO Participants}.
Different IO participant types (human operatives, grassroots campaigners, unwitting targets, trolls, bots) should be treated differently. We conjecture that features emerging from the above VPs (e.g., number of daily hashtags promoted, number of Patria apps installed, number and amplitude of activity spikes, counts of disinformation posted/promoted), could be used to train models to detect and even classify influence operations participants.

Previous work suggests promise for such an approach. Saeed et al.~\cite{SABDZS22} found that Reddit trolls differ from regular users, e.g., in their loose coordination activities. They leveraged these differences to develop relevant features and train a Reddit troll-detection model. Volkova and Bell~\cite{VB16} extracted features and trained a model to detect accounts involved in the 2014 Ukraine-Russia conflict, that were deleted by Twitter. Luceri et al.'s~\cite{LGF20} inverse reinforcement learning-based discovery that Russian Twitter trolls differ in their behavior when engaged by others or when their content is re-shared, further suggests potential to generalize previous work to similarly identify other types of IO participants.

\noindent
{\bf Monitor, Don't Censor}.
Study participants revealed strategies to groom backup accounts to address account suspensions, and also techniques to study shadowbanning. Instead, monitoring the activities of detected accounts would allow social networks to identify IO strategy shifts in real time, and implement subtler mechanisms to reduce the reach and impact of influence operations, and even turn them into echo chambers. We suggest two directions:

$\bullet$
{\bf Nudge Unwitting Participants}.
Social networks could deploy interventions to nudge unwitting targets of influence operations ($\S$~\ref{sec:background}) toward safer behaviors. This includes extending the social network client to signal to detected unwitting participants that certain accounts they follow have inauthentic behaviors, and suggest unfollowing such accounts. Further, signal when viewed posts are suspected of being promoted by influence operations, and suggest avoiding engagement. Keiser et al.~\cite{KWLLMM21} provide evidence that disinformation warnings that interrupt the user and
require interaction can inform and guide user behaviors.

$\bullet$
{\bf One Device One Vote}.
For Venezuelan operations, the difficulty to access mobile devices
%
%($\S$~\ref{appendix:findings:limitations}),
%
can be used to thwart attempts by operatives to game the system, e.g., the search rank of hashtags. For instance, Twitter's hashtag ranking algorithm could reduce the weight of contributions from devices associated with influence operations, e.g., to ensure that each device can contribute at most one vote per hashtag.

\section{Discussion and Limitations}
\label{sec:discussion}

\noindent
{\bf Pro-government vs. Opposition IOs}.
Our analysis reveals a complex dynamic between pro-government and opposition participants, that have different motivations, organizations, technical ecosystems, adversaries, and strategies. We also found common goals that include (1) preventing the suspension of their accounts, (2) ensuring their continued access to Twitter and their follower base after an account suspension, (3) acquiring many followers, from diverse groups, (4) receiving engagement, (5) promoting key hashtags to trending status, and (6) creating and distributing content that challenges the version of events of opposing factions.

%Our analysis reveals a complex dynamic between pro-government and opposition participants, that have different (1) motivations: paid/coerced operatives vs. true believers, (2) organizations: hierarchical vs. decentralized, (3) technical ecosystems, i.e., the Patria system, (4) adversaries: Twitter vs. government operatives, and (5) strategies: rate limiting efforts vs. avoiding suspension wars.

\noindent
{\bf Sockpuppets vs. Real Users}.
While a few participants rely on sockpuppet accounts, confirming previous findings~\cite{OC19, KCLS17, BloombergGuide}, we found many participants who only claim to control a few personal accounts. This supports recent Facebook reports that campaigns are starting to recruit real people into their amplification operations~\cite{FacebookIO}.

\noindent
{\bf IO = Collaborative Work}.
Our work confirms that influence operations are collaborative work~\cite{SAW19, FacebookIO, OC19, BloombergGuide}, whether through hierarchical structures consistent with previous reports~\cite{BloombergDocument, OC19}, or through flexible, decentralized structures. The communications technologies that form the basis of both structure types point at organizations of both pro-government and opposition participants~\cite{PN09}. We reveal a social network influence war between participants supporting opposing sides, where each side seeks to increase its influence and reach, and thwart the opponent's efforts.

%Participants claimed and we verified interest in international politics, including for the US and other Spanish-speaking countries. Fourteen participants expressed willingness to participate in campaigns for payment, including to target Spanish-speaking communities in the US. We provide evidence of participants contributing to discussions on controversial topics and promoting views from known fake news sites. This reveals a perhaps yet untapped potential for deception in US politics using imported workers.

\noindent
{\bf Effectiveness and Loopholes of Twitter Defenses}.
We observe that Twitter's defenses are effective to a certain degree to degrade the efforts of influence operations ($\S$~\ref{sec:results:defenses}).
%
%For instance, participant reported efforts by Twitter to prevent creation of accounts from flagged devices can be particularly effective against participants that have few resources. Twitter's strong enforcement of appropriate language has paid off, leading to participants claiming even self-censorship to eliminate inappropriate language. Twitter's (denied) use of shadowbanning may also frustrate the efforts of some participants to acquire fake engagement for their posts.
%
However, we also found that participants are pushing their activities beyond the limit admissible by Twitter, experiment with its defenses, learn from its responses, and share their lessons ($\S$~\ref{sec:results:defenses} and $\S$~\ref{sec:results:avoidance}). Nevertheless, despite finding that participants have posting strategies that successfully evade Twitter detection, we also identified behaviors that continue to render them vulnerable to detection ($\S$~\ref{sec:assurance}).

\noindent
{\bf Impact of Findings. Ethical Considerations Revisited}.
The goal of this study is not to help social networks detect and punish operatives. Instead, we leverage our findings to propose a shift from censoring to monitoring operatives, and detecting and nudging their unwitting targets. We believe that human participants will continue to play an important role in influence operations. However, the approach we propose in $\S$~\ref{sec:assurance:defenses} may reduce their perception of heavy-handed social network interference: operatives would no longer lose followers due to automatic suspensions, but only when their unwitting targets make an explicit effort to unfollow them. Their posts would no longer be shadowbanned, but may organically experience a reduced engagement from unwitting targets.

Giving users more information and control may thus also improve the public's trust in social networks. It may also indirectly nudge operatives to post higher quality, more trustworthy content, to avoid alienating their followers.

Our approach may also lead to strategy changes for the IO command and control ($\S$~\ref{sec:background:adversary}). For instance, they could provide more resources to participants (e.g., devices), and reduce the number of posts required by the Patria system to provide activity awards (now at 50 tweets and 300 retweets/day). This may reduce the detectability of operative accounts. They could also use better bots, that leverage, e.g., generative adversarial networks, to emulate human behaviors.

\noindent
{\bf Limitations}.
Our recruitment process was biased due to only contacting active Twitter users whose accounts were not bots, had at least 500 followers, were open to DMs from our accounts, read our DMs, and consented to our terms. We also observe the unexpected high mean participant age (50). This is perhaps due to older people having more time and willingness to discuss their experiences, among the accounts that we reached. We cannot claim that this age distribution applies to all people who post political content in Venezuela. We also note that our study focused on Venezuela, but cannot provide a complete picture of influence campaigns in Venezuela. Further, our findings do not apply to operations in other countries.

However, we present results from the first qualitative study conducted from the perspective of participants in influence operations, on their experiences and challenges associated with their online activities.

The quantitative analysis-based validation of participant claims is limited since we cannot identify all the accounts they control. Thus, our analysis can only validate a subset of the claims made by participants. This limitation applies to participants that claimed ownership of multiple, sockpuppet accounts, but are less obvious for participants that claimed only backup accounts. We confirm that all the accessible accounts revealed by interview participants were linked to their owner's identity.

\section{Conclusions}

In this paper we have reported findings from interviews with 19 influence operation participants that targeted Venezuela, and a quantitative investigation with data collected from participant-controlled accounts. We found that pro-government and opposition participants use similar content-promotion strategies, but have marked differences in their motivation, organization, technical solutions, adversaries, and detection avoidance strategies. Our findings complement previous work, suggesting a strategy adjustment for influence operations in Venezuela. We reveal however vulnerabilities of existing influence operations strategies, and suggest detection solutions.

\section{Acknowledgments}

This research was supported by NSF grants CNS-2013671 and CNS-2114911, and CRDF grant G-202105-67826. This publication is based on work supported by a grant from the U.S. Civilian Research \& Development Foundation (CRDF Global). Any opinions, findings and conclusions or recommendations expressed in this material are those of the author(s) and do not necessarily reflect the views of CRDF.

\bibliographystyle{plain}
\bibliography{bogdan,politicalspam,fakenews,socialfraud,theory}

\appendix

\end{document}